%% file: main.tex
\documentclass[aps,superscriptaddress,prx,twocolumn,10pt,floatfix,longbibliography]{revtex4-2}

\makeatletter\if@twocolumn\PassOptionsToPackage{switch}{lineno}\else\fi\makeatother

\usepackage{tabulary,graphicx,amsfonts,amsmath,amssymb,amsbsy,dcolumn,bm}
\usepackage[utf8]{inputenc}
\usepackage[T1]{fontenc}
\usepackage[whole]{bxcjkjatype}
\usepackage{adjustbox}
\usepackage{algorithm}
\usepackage{algorithmic}
\usepackage{braket}
\usepackage{comment}
\usepackage{natbib}
\usepackage{tikz}
\usepackage{xcolor}
\usepackage{xspace}

\usetikzlibrary{quantikz2}

\makeatletter
\let\MYcaption\@makecaption
\makeatother

\usepackage[font=footnotesize]{subcaption}

\makeatletter
\let\@makecaption\MYcaption
\makeatother

\usepackage{hyperref}
\hypersetup{citecolor=blue}

\usepackage{url,multirow,morefloats,floatflt,cancel,tfrupee}
\makeatletter

\AtBeginDocument{\@ifpackageloaded{textcomp}{}{\usepackage{textcomp}}}
\makeatother
\usepackage{colortbl}
\usepackage{xcolor}
\usepackage{pifont}
\usepackage[nointegrals]{wasysym}
\makeatletter

\def\mcWidth#1{\csname TY@F#1\endcsname+\tabcolsep}

\def\cAlignHack{\rightskip\@flushglue\leftskip\@flushglue\parindent\z@\parfillskip\z@skip}
\def\rAlignHack{\rightskip\z@skip\leftskip\@flushglue \parindent\z@\parfillskip\z@skip}

\@ifundefined{etal}{}{}

\usepackage{ifxetex}
\ifxetex\else\if@twocolumn\@ifpackageloaded{stfloats}{}{\usepackage{dblfloatfix}}\fi\fi

\AtBeginDocument{
\expandafter\ifx\csname eqalign\endcsname\relax
\def\eqalign#1{\null\vcenter{\def\\{\cr}\openup\jot\m@th
  \ialign{\strut$\displaystyle{##}$\hfil&$\displaystyle{{}##}$\hfil
      \crcr#1\crcr}}\,}
\fi
}

\AtBeginDocument{%
  \@ifpackageloaded{endfloat}%
   {\renewcommand\efloat@iwrite[1]{\immediate\expandafter\protected@write\csname efloat@post#1\endcsname{}}}{\newif\ifefloat@tables}%
}%

\def\BreakURLText#1{\@tfor\brk@tempa:=#1\do{\brk@tempa\hskip0pt}}
\let\lt=<
\let\gt=>
\def\processVert{\ifmmode|\else\textbar\fi}

\@ifundefined{subparagraph}{
\def\subparagraph{\@startsection{paragraph}{5}{2\parindent}{0ex plus 0.1ex minus 0.1ex}%
{0ex}{\normalfont\small\itshape}}%
}{}

\newcommand\role[1]{\unskip}
\newcommand\aucollab[1]{\unskip}
  
\@ifundefined{tsGraphicsScaleX}{\gdef\tsGraphicsScaleX{1}}{}
\@ifundefined{tsGraphicsScaleY}{\gdef\tsGraphicsScaleY{.9}}{}
\def\checkGraphicsWidth{\ifdim\Gin@nat@width>\linewidth
	\tsGraphicsScaleX\linewidth\else\Gin@nat@width\fi}

\def\checkGraphicsHeight{\ifdim\Gin@nat@height>.9\textheight
	\tsGraphicsScaleY\textheight\else\Gin@nat@height\fi}

\def\fixFloatSize#1{}%\@ifundefined{processdelayedfloats}{\setbox0=\hbox{\includegraphics{#1}}\ifnum\wd0<\columnwidth\relax\renewenvironment{figure*}{\begin{figure}}{\end{figure}}\fi}{}}
\let\ts@includegraphics\includegraphics

\def\inlinegraphic[#1]#2{{\edef\@tempa{#1}\edef\baseline@shift{\ifx\@tempa\@empty0\else#1\fi}\edef\tempZ{\the\numexpr(\numexpr(\baseline@shift*\f@size/100))}\protect\raisebox{\tempZ pt}{\ts@includegraphics{#2}}}}

\AtBeginDocument{\def\includegraphics{\@ifnextchar[{\ts@includegraphics}{\ts@includegraphics[width=\checkGraphicsWidth,height=\checkGraphicsHeight,keepaspectratio]}}}

\DeclareMathAlphabet{\mathpzc}{OT1}{pzc}{m}{it}

\def\URL#1#2{\@ifundefined{href}{#2}{\href{#1}{#2}}}

\def\UrlOrds{\do\*\do\-\do\~\do\'\do\"\do\-}%
\g@addto@macro{\UrlBreaks}{\UrlOrds}

\edef\fntEncoding{\f@encoding}

\makeatother

\def\style#1#2{#2}

\newif\ifmultipleabstract\multipleabstractfalse%
\newcommand{\up}[2]{#1^{(#2)}}
\newcommand{\upj}[1]{\up{#1}{j}}
\newcommand{\half}[1][1]{\frac{#1}{2}}
\newcommand{\quarter}[1][1]{\frac{#1}{4}}
\newcommand{\X}[1]{\mathcal{X}_{#1}}
\newcommand{\SX}[1]{\sqrt{\X{#1}}}
\newcommand{\Xplus}{\X{+}}
\newcommand{\Xminus}{\X{-}}
\newcommand{\Xpm}{\X{\pm}}
\newcommand{\Xmp}{\X{\mp}}
\newcommand{\sub}[2]{#1_{\mathrm{#2}}}
\newcommand{\Usub}[1]{\sub{U}{\mathrm{#1}}}
\newcommand{\Wcr}[1]{W_{#1}}
\newcommand{\Pop}[2]{P_{#1}\left(#2\right)}
\newcommand{\Rop}[2]{R_{#1}\left(#2\right)}

\newcommand{\ketbra}[2]{\ket{#1}\!\bra{#2}}
\newcommand{\gencx}{\mathrm{C}^1 \! X}
\newcommand{\gencz}{\mathrm{C}^1 \! Z}
\newcommand{\expfn}[1]{\exp\left( #1 \right)}
\newcommand{\qcone}{$Q_{c1}$\xspace}
\newcommand{\qctwo}{$Q_{c2}$\xspace}
\newcommand{\qt}{$Q_{t}$\xspace}
\newcommand{\us}{$\mu\mathrm{s}$\xspace}
\newcommand{\ibmdev}[1]{\texttt{ibm{\textunderscore}#1}}

\begin{document}

\title{Qudit-Generalization of the Qubit Echo and Its Application to a Qutrit-Based Toffoli Gate}

\author{Yutaro Iiyama}
\email{iiyama@icepp.s.u-tokyo.ac.jp}
\affiliation{International Center for Elementary Particle Physics (ICEPP), The University of Tokyo, 7-3-1 Hongo, Bunkyo-ku, Tokyo 113-0033, Japan}
\author{Wonho Jang}
\affiliation{Department of Physics, Graduate School of Science\unskip, The University of Tokyo\unskip, 7-3-1 Hongo\unskip, Bunkyo-ku\unskip, Tokyo 113-0033\unskip, Japan}
\author{Naoki Kanazawa}
\affiliation{IBM Quantum, IBM Research Tokyo, Tokyo, 103-8510, Japan}
\author{Ryu Sawada}
\affiliation{International Center for Elementary Particle Physics (ICEPP), The University of Tokyo, 7-3-1 Hongo, Bunkyo-ku, Tokyo 113-0033, Japan}
\author{Tamiya Onodera}
\affiliation{IBM Quantum, IBM Research Tokyo, Tokyo, 103-8510, Japan}
\author{Koji Terashi}
\affiliation{International Center for Elementary Particle Physics (ICEPP), The University of Tokyo, 7-3-1 Hongo, Bunkyo-ku, Tokyo 113-0033, Japan}

\date{\today}

\begin{abstract}
\input{abstract}
\end{abstract}\def\keywordstitle{Keywords}

\maketitle

\section{Introduction}\label{sec:intro}
\input{intro}

\section{Basis cycling}\label{sec:basis_cycling}
\input{basis_cycling}

\section{Toffoli and CCZ gate decomposition using a qutrit}\label{sec:toffoli}
\input{toffoli}

\section{Conclusion}\label{sec:discussion}
\input{discussion}

\textit{Note added.} Recently, we became aware of a related work that provides a general formulation of dynamical decoupling for qudits that is practically similar to what is described in this paper~\cite{Tripathi2024-gb}.

The source code and usage examples of our qutrit calibration programs are publicly available~\cite{github-qutrit-experiments}.

\section*{Acknowledgments}
\input{acknowledgments}

YI and WJ contributed equally to this work.

\appendix

\section{Qudit phase accumulation and control}\label{app:qudit_phase}
\input{qudit_phase}

\section{Reframing qubit echo sequences}\label{app:reframing_echo}
\input{reframing_echo}

\section{Cross resonance effective Hamiltonian with a qudit control}\label{app:qudit_hcr}
\input{qudit_hcr}

\section{Gate Calibration}\label{app:calibration}
\input{calibration}

\section{Quantum channel error analysis}\label{app:error_analysis}
\input{error_analysis}

\section{Qubit-based Toffoli decomposition}\label{app:qubit_toffoli}
\input{qubit_toffoli}

\bibliography{references}

\end{document}

%% file: abstract.tex
The fidelity of certain gates on noisy quantum computers may be improved when they are implemented using more than two levels of the involved transmons. The main impediments to achieving this potential are the dynamic gate phase errors that cannot be corrected via calibration. The standard tool for countering such phase errors in two-level qubits is the echo protocol, often referred to as the dynamical decoupling sequence, where the evolution of a qubit is punctuated by an even number of X gates. We introduce basis cycling, which is a direct generalization of the qubit echo to general qudits, and provide an analytic framework for designing gate sequences to produce desired effects using this technique. We then apply basis cycling to a Toffoli gate decomposition incorporating a qutrit and obtain CCZ gate fidelity values up to 93.8$\pm$0.1\%, measured by quantum process tomography, on IBM quantum computers. The gate fidelity remains stable without recalibration even while the resonant frequency of the qutrit fluctuates, highlighting the dynamical nature of phase error cancellation through basis cycling. Our results demonstrate that one of the biggest difficulties in implementing qudit-based gate decompositions on superconducting quantum computers can be systematically overcome when certain conditions are met, and thus open a path toward fulfilling the promise of qudits as circuit optimization agents.

%% file: intro.tex
Quantum computers utilizing computational units with more than two levels (qudits) are natural extensions to the conventional qubit-based computers. While qudits are expected to have substantially enhanced capabilities for data encoding and computation compared to qubits~\cite{Parasa2011-yu, Campbell2014-cf, Gokhale2019-wi, Wang2020-ny, Pavlidis2021-jd, Gustafson2021-zo, Deller2023-nw}, the potential of fully qudit-based quantum circuits and qudit-native algorithms remains largely unexplored. Instead, there has recently been a growing interest across the community in the local and transient use of qudits, specifically the three-level variants (qutrits), within otherwise qubit-based circuits to improve their efficiency and accuracy~\cite{Ralph2007-pu, Lanyon2008-ys, fedorov_implementation_2012, Egger2018-vg, Magnard2018-wy, Baekkegaard2019-mj, baker2020efficient, 2108.01652, Elder2020-ch, Nikolaeva2024-sr}.
%perhaps driven by the experimental reality that qutrits tend to be coherently controllable for a much shorter time scale than for qubits.

%In a transient application, qutrit gates are embedded at critical points of otherwise qubit-only circuits, and are used for example as temporary information storage~\cite{Baker2020-rb,2108.01652}; modifier of controlled operations~\cite{Lanyon2008-ys,Asymptotic,fedorov_implementation_2012}; protection against qubit decoherence~\cite{Elder2020-ch}; and a conduit for fast qubit reset~\cite{Egger2018-vg,Magnard2018-wy}. Unless immediately followed by a measurement, as in the latter two examples, the state space expansion is ``uncomputed'', i.e., the state vector is rotated back into the qubit space by the last qutrit gate.

The transmon~\cite{Koch2007-qv}, which is one of the most established platforms for implementing a qubit, is also readily usable as a qutrit. A qubit in a transmon is represented by its ground ($\ket{0}$) and first excited ($\ket{1}$) states, where polar rotations around the Bloch sphere are induced by microwave drive signals tuned to the excitation energy. A similar two-level structure is formed with $\ket{1}$ and the second excited state ($\ket{2}$), with a distinct drive frequency thanks to the anharmonicity of the transmon. An arbitrary state in the three-dimensional Hilbert space spanned by $\{\ket{0}, \ket{1}, \ket{2}\}$ can be realized by combining the control signals for the two binary systems.

A major obstacle in utilizing the $\ket{2}$ state of the transmon is the high charge-noise sensitivity of the state energy, resulting from its charge dispersion that is typically an order of magnitude larger than that of the $\ket{1}$ state~\cite{Koch2007-qv}. Because charge noise is prevalent on existing transmon devices, the dephasing time of the $\ket{1}$-$\ket{2}$ system is typically comparable to the duration of a shallow circuit~\cite{Siddiqi}. This problem of fast decoherence appears to be the main motivation for restricting the qutrit usage to be transient, i.e., to exploit the full qutrit state space only within a short gate sequence. However, the effect of inhomogeneous broadening caused by low-frequency charge noise still remains, manifesting itself as coherent but uncontrollable phase errors in qutrit gates.

Coherent phase errors are not specific to qutrit gates. For instance, qubit resonant frequency can also shift over time. On statically coupled fixed-frequency transmons such as those found on devices provided by IBM Quantum, more prominent effects arise from the AC Stark shift caused by cross resonance (CR) driving and the cross-Kerr interaction between neighboring qubits~\cite{rotary_echo, Ku2020-vh}. Remarkably, countermeasures to all three effects, namely refocusing~\cite{Hahn1950-ju}, local-term cancellation~\cite{Corcoles2013-wg}, and dynamical decoupling~\cite{Faoro2004-cd, Pokharel2018-bv, Jurcevic2021-ox}, can be implemented with a common technique, the echo sequence, in which the qubit evolution is punctuated by an even number of X pulses to induce a self-cancelling phase accumulation. A similar robust and effective protocol for qutrits is desirable, since qutrits are even more strongly affected by all of these phase errors than qubits are. Dynamical decoupling of qutrits and refocusing of the $\ket{1}$-$\ket{2}$ system are discussed in Refs. \cite{Siddiqi} and \cite{galda2021implementing}, respectively, but their implementations are purpose-specific and leave unclear how the echo sequence might generalize to qutrits, let alone general qudits.

In this paper, we give a reinterpretation to the operational principle of the echo sequence as a cyclic permutation of the computational basis. A qudit generalization follows immediately from this perspective. The generalized echo protocol, which we call \text{basis cycling}, is in fact a formalization of the idea that appears in Ref. \cite{Siddiqi} as a purpose-specific gate sequence. We clarify the underlying mechanism and the applicability condition of the technique. Local phase cancellation and qutrit refocusing based on basis cycling are then demonstrated in Toffoli and CCZ gate decompositions equivalent to the one presented in Ref.~\cite{Ralph2007-pu}.

The Toffoli gate is a three-qubit entangling gate that enables arbitrary boolean operations over a quantum register~\cite{Toffoli1980-fg}, and when combined with the Hadamard gate, forms a universal gate set~\cite{Shi2003-lf}. It is also an essential component of multi-controlled operations and arithmetic circuits such as adders~\cite{Cuccaro2004-oo}, both of which in turn constitute critical subroutines for various important quantum algorithms~\cite{Haner2017-yz,Grover}. Furthermore, certain error correction protocols depend on this gate~\cite{Cory_1998,Schindler2011-fh,qec_3}. In spite of its significance and ubiquity, the Toffoli gate and the locally equivalent CCZ gate are difficult to implement with infidelity in the $10^{-3}$ scale or lower, desirable for repeated use in practical quantum computation tasks, on today's general-purpose superconducting quantum computers. The main reason for the difficulty is the need to decompose all three-qubit gates into one- and two-qubit gates that are natively supported on the given platform. The most efficient decomposition of the Toffoli gate over linearly connected qubits requires eight two-qubit gates~\cite{Duckering2021-xh, MQ_Cruz2024-at}, each of which carry small but non-negligible error. Meanwhile, it has been shown that the required number of entangling gates is significantly reduced when qudits are utilized in the decomposition~\cite{Ralph2007-pu, Lanyon2008-ys, fedorov_implementation_2012, 2108.01652, qec_3, Baekkegaard2019-mj, galda2021implementing, Cervera-Lierta2022-ds, Nikolaeva2022-jk}, giving a strong motivation for the research of robust qutrit control technologies.

In fact, multiple successful implementations of the Toffoli and CCZ gates using qudits, particularly qutrits, on superconducting devices exist~\cite{fedorov_implementation_2012, qec_3, galda2021implementing, 2108.01652}, with the most recent work~\cite{Nguyen2024-eo} reporting a CCZ gate fidelity as high as 96.18$\pm$0.05\%, measured via cycle benchmarking~\cite{Erhard2019-uc}. This result is comparable to the best fidelity achieved with the qubit-only decomposition on an IBM device. The gate fidelity values obtained in our demonstration are slightly lower, but are manifestly stable beyond the drift time scale of the $\ket{1}$-$\ket{2}$ resonance frequency thanks to 
the dynamical cancellation of phase errors.

Our contributions in this paper are thus twofold: To formulate the concept of basis cycling and introduce its applications; and to present a novel Toffoli gate decomposition using a basis-cycled qutrit. We employ Qiskit Pulse~\cite{Qiskitpulse} to implement qutrit gates on IBM Quantum computers. The automated gate calibration program is written within the Qiskit Experiments framework~\cite{Kanazawa2023-tz} and is made publicly available, allowing anyone with necessary access to appropriate interfaces to construct high-precision qutrit gates that can be used for any purpose.

The remainder of this paper is organized as follows. In Section \ref{sec:basis_cycling} we define basis cycling and describe two circuit optimization techniques based on this concept. These techniques are put into practice in Section \ref{sec:toffoli} where we present the Toffoli gate decomposition and report the experimental results of its implementation. The conclusion is given in Section \ref{sec:discussion}.

%% file: basis_cycling.tex
\subsection{Definition}
\label{subsec:bc_definition}

To formulate basis cycling, we first study the echo sequence on a system comprising multiple qubits. In the following, we call the qubit onto which the echo pulses are applied as the main qubit, and the remaining qubits as the environment. For ease of discussion, we consider the so-called asymmetric form of the echo sequence, which can be expressed in terms of unitary operators as
\begin{equation}\label{eqn:qubit_echo_unitaries}
U = X \up{V}{1} X \up{V}{0}
\end{equation}
The operators act on the system from right to left. Here, $X$ represents the $X$ gate on the main qubit, with the identity operation on the environment implied. The unitaries $\upj{V}$ are diagonal with respect to the main qubit.

While such echo sequences are typically analyzed in terms of qubit unitary operators, here we use a representation with projectors onto the computational states of the main qubit, as this form helps clarify the generalization of qubit echo into basis cycling. With the projectors, the unitaries $\upj{V}$ are expressed as
\begin{equation}\label{eqn:qubit_vj_decomposition}
  \upj{V} = e^{i\upj{\phi}_0} \proj{0} \otimes \upj{v}_0 + e^{i\upj{\phi}_1} \proj{1} \otimes \upj{v}_1,
\end{equation}
where $\upj{\phi}_k$ are real numbers and $\upj{v}_k$ are unit-determinant unitaries acting on the environment. The projectors are swapped when the $\upj{V}$ operator is interposed between two $X$s as
\begin{equation}
  X \up{V}{1} X = e^{i\up{\phi}{1}_0} \proj{1} \otimes \up{v}{1}_0 + e^{i\up{\phi}{1}_1} \proj{0} \otimes \up{v}{1}_1.
\end{equation}
Therefore, the overall echo sequence yields
\begin{multline}\label{eqn:qubit_basis_cycling}
  U = e^{i \left(\up{\phi}{1}_1 + \up{\phi}{0}_0\right)} \proj{0} \otimes \up{v}{1}_1 \up{v}{0}_0 \\
    + e^{i \left(\up{\phi}{1}_0 + \up{\phi}{0}_1\right)} \proj{1} \otimes \up{v}{1}_0 \up{v}{0}_1.
\end{multline}

All of the circuit optimization techniques involving the echo sequence exploit certain relations among the phase factors $\upj{\phi}_k$ and unitaries $\upj{v}_k$ to engineer the desired effects. For example, if $\up{v}{1}_1 \up{v}{0}_0 = \up{v}{1}_0 \up{v}{0}_1$, the main qubit is decoupled from the environment, whereas if $\up{\phi}{1}_1 + \up{\phi}{0}_0 = \up{\phi}{1}_0 + \up{\phi}{0}_1$, the local phase of the main qubit is factored out as global. Analyses of qubit refocusing, dynamical decoupling, and echoed cross resonance (ECR) in this formulation are presented in Appendix~\ref{app:reframing_echo}.

Equation \eqref{eqn:qubit_basis_cycling} can be interpreted as each computational basis of the main qubit cyclically collecting the phase factors and the environment unitaries. From this perspective, we can identify a phase-cancelling sequence for a $d$-level qudit, which we call basis cycling, as
\begin{equation}\label{eqn:qudit_basis_cycling_1}
  U = \prod_{j=0}^{d-1} \Xplus \upj{\mathcal{V}}, \\
\end{equation}
where $\Xplus = \sum_{l=0}^{d-1} \ket{l+1}\bra{l}$ (with $\ket{d}=\ket{0}$) is the cyclic level-shift operator acting on the main qudit that generalizes the $X$ operator, and $\upj{\mathcal{V}}$ are the unitary operators diagonal with respect to this qudit. The $\Xplus$ gate is implemented with a chain of ``$\pi$ pulses'' (full level transitions) in successive two-level spaces. Hereafter, the product symbol signifies left multiplication of the operators. In the projector representation, $\upj{\mathcal{V}}$ can be given a decomposition similar to equation \eqref{eqn:qubit_vj_decomposition}, and we have
\begin{equation}\label{eqn:qudit_basis_cycling_2}
  U = \sum_{l=0}^{d-1} \left[ e^{i \sum_{j=0}^{d-1} \up{\phi}{j}_{l+j}} \proj{l} \otimes \prod_{j=0}^{d-1} \up{v}{j}_{l+j} \right].
\end{equation}
Subscripts of $\phi$ and $v$ should be understood with modulo $d$.

It should be noted that $U$ is itself diagonal with respect to the main qudit by virtue of $\upj{\mathcal{V}}$ operators being diagonal. Equation~\eqref{eqn:qudit_basis_cycling_1} thus elucidates the applicability condition of basis cycling as well as defines it; just like qubit echo, basis cycling is valid when there is a diagonal operation that is replaceable with a product of diagonal and level-shift operators.

Basis cycling can realize various effects depending on the relations of the $\upj{\phi}_k$ factors and $\upj{v}_k$ operators. In the following, we give explicit calculations for qutrit refocusing and cyclic cross resonance.

\subsection{Qutrit refocusing}
\label{subsec:qutrit_refocusing}

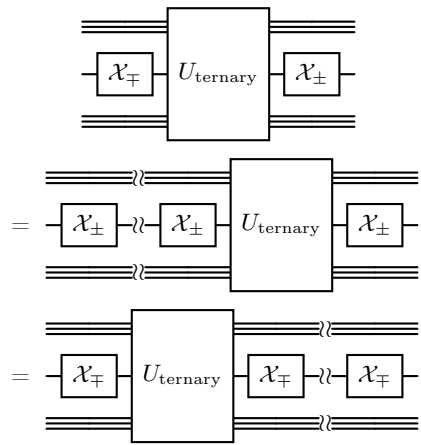
\begin{figure}
  \begin{quantikz}[wire types={b,q,b},column sep=0.2cm,row sep=0.1cm,classical gap=0.07cm]
      &                          & \gate[3,disable auto height]{U_{\mathrm{ternary}}} &                          & \\
      & \gate{\mathcal{X}_{\mp}} &                                                    & \gate{\mathcal{X}_{\pm}} & \\
      &                          &                                                    &                          &
  \end{quantikz} \\
  =
  \begin{quantikz}[wire types={b,q,b},column sep=0.2cm,row sep=0.1cm,classical gap=0.07cm]
     &                           & \wr\wr &                          & \gate[3,disable auto height]{U_{\mathrm{ternary}}} &                          & \\
     & \gate{\mathcal{X}_{\pm}}  & \wr\wr & \gate{\mathcal{X}_{\pm}} &                                                    & \gate{\mathcal{X}_{\pm}} & \\
     &                           & \wr\wr &                          &                                                    &                          &
  \end{quantikz} \\
  =
  \begin{quantikz}[wire types={b,q,b},column sep=0.2cm,row sep=0.1cm,classical gap=0.07cm]
     &                          & \gate[3,disable auto height]{U_{\mathrm{ternary}}} &                          & \wr\wr &                           & \\
     & \gate{\mathcal{X}_{\mp}} &                                                    & \gate{\mathcal{X}_{\mp}} & \wr\wr & \gate{\mathcal{X}_{\mp}}  & \\
     &                          &                                                    &                          & \wr\wr &                           &
  \end{quantikz}
  \caption{Standard form of a transient ternary circuit and qutrit refocusing sequences.}
  \label{fig:u_ternary_circuit}
\end{figure}

By definition, the part of a circuit that transiently ``upgrades'' a qubit to a qutrit is bounded by the first and last $\ket{1}$-$\ket{2}$ space gates. Using gate identities, such a sub-circuit can be given a standard form in the top row of Figure~\ref{fig:u_ternary_circuit}, where $\Xminus = \Xplus^{\dagger} = \sum_{l=0}^{2} \ket{l}\bra{l + 1}$. The corresponding unitary $V$ is
\begin{equation}\label{eqn:general_ternary}
  V = \Xpm \Usub{ternary} \Xmp.
\end{equation}
For a qutrit, the cyclic shift operators are implemented as
\begin{align}
  \Xplus = \X{0} \X{1}, \\
  \Xminus = \X{1} \X{0},
\end{align}
where $\X{0}$ and $\X{1}$ are the standard $\pi$ pulse gates in the $\ket{0}$-$\ket{1}$ and $\ket{1}$-$\ket{2}$ spaces, respectively. Using the generators
\begin{equation}
  X_{k} = \ket{k}\bra{k+1} + \ket{k+1}\bra{k},
\end{equation}
the $\pi$ pulse gates are expressed as
\begin{equation}
  \begin{split}
  \X{0} & = \Pop{2}{-\half[\pi]} \expfn{-\half[i \pi] X_0} \\
        & = -i (\ketbra{0}{1} + \ketbra{1}{0} + \proj{2})
  \end{split}
\end{equation}
and
\begin{equation}
  \begin{split}
  \X{1} & = \Pop{0}{-\half[\pi]} \expfn{-\half[i \pi] X_1} \\
        & = -i (\proj{0} + \ketbra{1}{2} + \ketbra{2}{1}).
  \end{split}
\end{equation}
The single-qutrit gates $P_l \; (l=0,1,2)$, inserted to correct for the so-called geometric phase, are defined as
\begin{equation}
  P_l(\phi) = \sum_{k \neq l} \proj{k} + e^{i\phi} \proj{l},
\end{equation}
and is described in detail in Appendix~\ref{app:qudit_phase}.

Due to the large charge dispersion of the $\ket{2}$ level of the transmon, the frame frequency for $\ket{1}$-$\ket{2}$ control is in practice always detuned from the true resonance by some time-varying value $\delta$. A $\pi$ pulse in this frame leads to a $\X{1}$ gate with a coherent but unknown phase shift $\delta t$, where $t$ is the time measured from the first $\X{1}$ gate in the overall circuit. See Appendix~\ref{app:qudit_phase} for derivation. Denoting the realized gate as
\begin{equation}
  \X{1}^{(t)} = \Pop{2}{-\delta t} \X{1} \Pop{2}{\delta t},
\end{equation}
the resulting cyclic shift gates are
\begin{align}
  \Xplus^{(t)} & = \Xplus \Pop{1}{-\delta t} \Pop{2}{\delta t},  \label{eqn:xplus_phase} \\
  \Xminus^{(t)} & = \Xminus \Pop{0}{\delta t} \Pop{1}{-\delta t}.
\end{align}

When $\Usub{ternary}$ in equation~\eqref{eqn:general_ternary} has a form
\begin{equation}\label{eqn:qutrit_diagonal_u}
  \Usub{ternary} = \sum_{l=0}^{2} e^{-i \delta \alpha_l} \proj{l} \otimes v_l,
\end{equation}
where the unitaries $v_l$ act on the remaining qubits and do not depend on $\delta$, this phase error from detuning can be factored out from $V$ as an overall $P_2$ via basis cycling. Because $V$ is a qubit-space unitary, the phase error is then effectively nullified. Coefficients $\alpha_l$ appear in equation~\eqref{eqn:qutrit_diagonal_u} because $\Usub{ternary}$ represents a circuit exploiting a qutrit, and therefore may involve $\X{1}$ and other $\ket{1}$-$\ket{2}$ space gates, which are affected by the detuning error. These coefficients can be calculated from timings of the qutrit gates within $\Usub{ternary}$.

To demonstrate this cancellation of the phase error, we analyze the case with upper signs in equation~\eqref{eqn:general_ternary}. Similar argument applies to the lower signs. From the identity $\Xplus^3 = \Xplus \Xminus = I$, the first 
$\Xminus$ in equation~\eqref{eqn:general_ternary} can be decomposed into two $\Xplus$ gates, as shown in the second row of Figure~\ref{fig:u_ternary_circuit}. Denoting the two intervals between the $\Xplus$ gates as $\tau_1$ and $\tau_2$, the modified $V$ is
\begin{equation}
\begin{split}
  V' = & \Xplus^{(t + \tau_1 + \tau_2)} \Usub{ternary} \Xplus^{(t + \tau_1)} \Xplus^{(t)} \\
   = & e^{i\delta (\tau_2 - \alpha_2)} \proj{0} \otimes v_2 \\
  & + e^{i\delta (\tau_1 - \alpha_0)} \proj{1} \otimes v_0 \\
  & + e^{-i\delta (\alpha_1 + \tau_1 + \tau_2)} \proj{2} \otimes v_1.
  \end{split}
\end{equation}
Therefore, if $\tau_1$ and $\tau_2$ are chosen so that
\begin{equation}
  \tau_2 - \alpha_2 = \tau_1 - \alpha_0,
\end{equation}
we see that up to a global phase
\begin{equation}
  V' = \Pop{2}{\delta (-\alpha_0 - \alpha_1 + 2\alpha_2 - 3\tau_2)} V,
\end{equation}
as anticipated.

% We note that in the special case when the ternary sub-circuit has a form
% \begin{equation}
%   V = \Xpm U_2 \Xpm U_1 \Xpm,
% \end{equation}
% where the unitaries $U_i$ are diagonal with respect to the qutrit, refocusing may take place without first transforming this to the standard form of equation~\eqref{eqn:general_ternary}. In fact, the CyCR sequence defined previously fits in this case and is actually free from detuning phase error.

\subsection{Cyclic cross resonance}
\label{subsec:cyclic_cross_resonance}

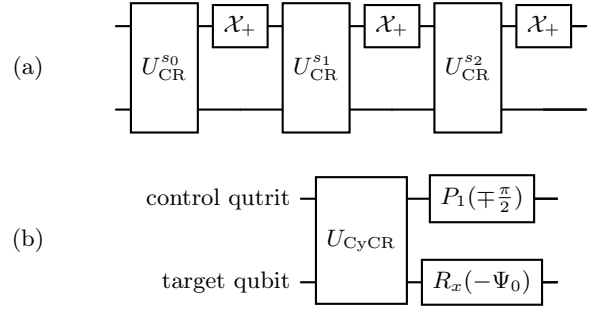
\begin{figure}
  \begin{subcaptiongroup}
    (a)
    \phantomcaption\label{fig:u_ccr}
    \begin{minipage}[t]{0.92\linewidth}
    \begin{quantikz}[column sep=0.2cm]
     & \gate[2]{\Usub{CR}^{s_0}} & \gate{\Xplus} & \gate[2]{\Usub{CR}^{s_1}} & \gate{\Xplus} & \gate[2]{\Usub{CR}^{s_2}} & \gate{\Xplus} & \qw \\
     &                           &               &                           &               &                           &               & \qw
    \end{quantikz}
    \end{minipage}
  \end{subcaptiongroup}
  
  \hfill\vspace{8pt}\\
  
  \begin{subcaptiongroup}
    (b)
    \phantomcaption\label{fig:generalized_cx_ccr}
    \begin{minipage}[t]{0.92\linewidth}
    \begin{quantikz}[column sep=0.2cm]
      \lstick{control qutrit} & \gate[2]{\Usub{CyCR}} & \gate{P_1(\mp \half[\pi])} & \qw \\
      \lstick{target qubit}   &                      & \gate{R_x(-\Psi_0)} & \qw
    \end{quantikz}
    \end{minipage}
  \end{subcaptiongroup}

  \caption{(a) Qutrit ($d=3$) cyclic cross resonance (CyCR). Superscripts $s_j$ represent the CR drive polarity of each pulse. In general, the $\Usub{CR}$-$\Xplus$ sequence is repeated $d$ times. (b) Generalized CX gate using CyCR. See Appendix~\ref{app:qudit_phase} for the definition of the $P_1$ gate.}
\end{figure}

Given a transversely coupled two-qudit system, the CR interaction occurs when a qudit (control) is driven at the $\ket{k}$-$\ket{k+1}$ transition frequency of the other qudit (target). The resulting effective Hamiltonian, expressed with operators on the control qudit ($d$ levels) on the left, is
\begin{equation}\label{eqn:qudit_hcr}
  H_{\mathrm{CR}}^{\pm} = \half \sum_{l=0}^{d-1} \proj{l} \otimes \left(\nu_{l} I \pm \omega_{l} X_{k} + \Delta_l\right).
\end{equation}
Here, $I$ is the identity operator, and $\Delta_l$ is a small traceless diagonal operator. The signs correspond to drive polarity, i.e., the oscillator phase of the drive signal being set to 0 or $\pi$. In terms of dominant physical sources, coefficients $\nu_l$ are manifestations of the AC Stark shift of the control qudit, $\omega_l$ represent the interaction between the static coupling and the CR drive, and $\Delta_l$ arise from the ``always-on'' interaction between the qudits. See Appendix~\ref{app:qudit_hcr} for the derivation of equation~\eqref{eqn:qudit_hcr}.

In a more familiar setup of a two-qubit system, we have $\Delta_l = \delta_l Z$, and equation~\eqref{eqn:qudit_hcr} reduces to
\begin{equation}\label{eqn:qubit_hcr}
  H_{\mathrm{CR}}^{\pm} = \pm\half[\nu_{IX}] IX + \half[\nu_{IZ}] IZ + \half[\nu_{ZI}] ZI \pm \half[\nu_{ZX}] ZX + \half[\nu_{ZZ}] ZZ,
\end{equation}
omitting the product symbols. The translations between the coefficients are given by
\begin{equation}
\begin{split}
  & \nu_{IX} = \half(\omega_0 + \omega_1), \;
  \nu_{IZ} = \half(\delta_0 + \delta_1), \;
  \nu_{ZI} = \half(\nu_0 - \nu_1), \\
  & \nu_{ZX} = \half(\omega_0 - \omega_1), \;
  \nu_{ZZ} = \half(\delta_0 - \delta_1).
\end{split}
\end{equation}

The utility of CR is in the $ZX$ term in equation~\eqref{eqn:qubit_hcr}, which serves as a basis for constructing the controlled-X (CX) gate~\cite{CR_1}. Unfortunately, a non-zero $ZX$ term is generally accompanied by non-zero $IX$ and $ZI$ terms, which spoil the CX gate composition with a single CR pulse. The ECR sequence was invented to eliminate the effect of these term by embedding two opposite-polarity CR pulses in an qubit echo sequence~\cite{tuning}. With the addition of the rotary tones~\cite{rotary_echo}, which suppress the effects from the $IZ$ and $ZZ$ terms, an almost-pure $ZX$ rotation can be achieved in two-qubit systems.

To realize a similar ``$\text{(diagonal)} \otimes X$''-type rotation in a general two-qudit system, ECR can be generalized to \textit{cyclic} cross resonance (CyCR). First, assuming for simplicity that $\Delta_l$ are small enough to be ignored or can be somehow eliminated, the unitary enacted by a CR pulse is written approximately as
\begin{equation}\label{eqn:ideal_cr_unitary}
  \Usub{CR}^{\pm} = \sum_{l=0}^{d-1} \proj{l} \otimes \expfn{-\half[i] [\phi_l I \pm \psi_l X_{k}]},
\end{equation}
with some coefficients $\phi_l$ and $\psi_l$. The signs of the $X_k$ term again represent the drive polarity. Then, repeating this unitary in a basis cycle sequence with the drive polarity of the $j$th pulse $s_j=\pm 1$, as shown in Figure~\ref{fig:u_ccr}, gives us the CyCR unitary
\begin{equation}\label{eqn:qudit_ccr}
\begin{split}
  \Usub{CyCR} = & \sum_{l=0}^{d-1} \proj{l} \otimes \expfn{-\half[i] \sum_{j=0}^{d-1} [\phi_{l+j} I + s_j \psi_{l+j} X_k]} \\
  = & e^{-\half[i] \Phi} \sum_{l=0}^{d-1} \proj{l} \otimes \expfn{-\half[i] \Psi_l X_k},
\end{split}
\end{equation}
where $\Phi = \sum_j \phi_j$ and $\Psi_l = \sum_j s_j \psi_{l+j}$. As intended, the $\phi_l$ factors are summed up into a global phase, and the overall unitary becomes an $l$-dependent $X_k$ rotation of the target qudit. Note that if all $s_j$ are equal, $\Psi_l$ will not depend on $l$, i.e., no entanglement is generated by $\Usub{CyCR}$.

Specifically, in a qutrit-qubit system, if the relation
\begin{equation}\label{eqn:ccr_cx_condition}
  \Psi_0 = \Psi_2 = \Psi_1 \pm \pi
\end{equation}
is satisfied, the CyCR unitary omitting the global phase becomes
\begin{equation}
\begin{split}
  \Usub{CyCR} = & (\proj{0} + \proj{2}) \otimes \Rop{x}{\Psi_0} \\
  & + \proj{1} \otimes \Rop{x}{\Psi_0 \mp \pi},
\end{split}
\end{equation}
where
\begin{equation}
  \Rop{x}{\theta} = \expfn{-\half[i\theta] X}
\end{equation}
is the standard $R_x$ gate on the target qubit. This $\Usub{CyCR}$ can be combined with local operations as depicted in Figure~\ref{fig:generalized_cx_ccr} to form the so-called generalized CX gate
\begin{equation}\label{eqn:generalized_cx}
  \gencx = (\proj{0} + \proj{2}) \otimes I + \proj{1} \otimes X.
\end{equation}
% Calibration of the CyCR gate, including how to realize the condition~\eqref{eqn:ccr_cx_condition} and how to deal with the $\Delta_l$ operators that are in fact non-negligible, is discussed in Appendix~\ref{app:calibration}.

%% file: toffoli.tex
\subsection{Overview}
\label{subsec:toffoli_overview}

\begin{figure*}
  \centering
  \begin{quantikz}[column sep=0.2cm]
     \lstick{\qcone} & \ctrl{2}   & \qw \\
     \lstick{\qctwo} & \control{} & \ghost{\Xminus} \qw \\
     \lstick{\qt}    & \targ{}    & \qw
    \end{quantikz}
    =
    \begin{quantikz}[column sep=0.2cm]
     &                & \ctrl{1} &                           & \ctrl{1} &               & \qw \\
     & \gate{\Xminus} & \targ{}  & \measure{1}\wire[d][1]{q} & \targ{}  & \gate{\Xplus} & \qw \\
     &                &          & \targ{}                   &          &               & \qw
    \end{quantikz}
    =
    \begin{quantikz}[column sep=0.2cm]
     &               & \wr\wr &               & \ctrl{1} &                           & \ctrl{1} &               & \qw \\
     & \gate{\Xplus} & \wr\wr & \gate{\Xplus} & \targ{}  & \measure{1}\wire[d][1]{q} & \targ{}  & \gate{\Xplus} & \qw \\
     &               & \wr\wr &               &          & \targ{}                   &          &               & \qw
  \end{quantikz}
  \caption{Ternary decompositions of the Toffoli gate without (center) and with (right) basis cycling.}
  \label{fig:toffoli_overview}
\end{figure*}
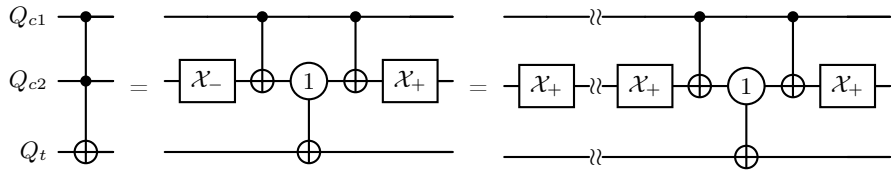

A Toffoli or CCZ gate on three qubits in a linear topology requires at least eight two-qubit gates, if the basis gate set used for the decomposition comprises single-qubit gates and the CX or a locally equivalent gate as the sole two-qubit gate~\cite{Duckering2021-xh, MQ_Cruz2024-at}. However, the number of entangling gates can be reduced to as low as three under a ternary decomposition~\cite{Nikolaeva2022-jk}, such as the one presented in Ref.~\cite{Ralph2007-pu} and recast in Figure~\ref{fig:toffoli_overview}. In fact, a multi-controlled X gate with $n$ ($\geq 2$) control qubits can be implemented using $2n-1$ entangling gates under a ternary decomposition, whereas qubit-only decompositions would require $6n-4$ such gates~\cite{Maslov2016-ad,Duckering2021-xh}. Using fewer entangling gates may result in a better gate fidelity, which would be a concrete benefit of using qutrits.

The minimum requirement for realizing the qutrit-based Toffoli gate is a high-quality implementation of the $\X{1}$ gate, which is attainable on the IBM platform using the pulse-controlling feature of Qiskit~\cite{Qiskitpulse}. In fact, through an originally developed calibration sequence documented in Appendix~\ref{app:calibration}, we realize $\X{1}$ gates on many different qubits with error per gate as low as $(2.2 \pm 0.5) \times 10^{-4}$ based on interleaved randomized benchmarking~\cite{Morvan2021-iy}.

Having a reliable $\X{1}$ gate is however not enough to achieve a high Toffoli gate fidelity, because ternary decompositions are affected by dephasing errors of multiple sources. As discussed in the previous sections, many of these errors can be countered using basis cycling techniques. Thus, to demonstrate the effectiveness of this approach, we implement the Toffoli gate under the decomposition in Figure~\ref{fig:toffoli_overview}, where \qcone, \qctwo, and \qt are the first control, second control, and target units. In this decomposition, \qctwo is internally operated as a qutrit, and the CX gate acting between this qutrit and \qt corresponds to $\gencx$ in equation~\eqref{eqn:generalized_cx}. To instead compose a CCZ gate, this $\gencx$ is replaced with a similarly defined generalized controlled-Z gate $\gencz$.

First, we find that the original decomposition of Figure~\ref{fig:toffoli_overview} center has the same form as Figure~\ref{fig:u_ternary_circuit}, and is therefore prone to phase errors from qutrit detuning $\delta$. To refocus the qutrit, the circuit needs to be transformed to a basis-cycled sequence of Figure~\ref{fig:toffoli_overview} right, and the central part involving three CX gates must have a form of $\Usub{ternary}$ in equation~\eqref{eqn:qutrit_diagonal_u}.

This latter requirement implies that the $\delta$-dependent factor in $\gencx$ must commute with the surrounding CX gates, i.e., the generalized CX gate with the detuning error is
\begin{equation}\label{eqn:gencx_prime}
  \gencx' = \Pop{2}{\delta \alpha} \gencx
\end{equation}
for some $\alpha$. Since we are implementing the Toffoli gate on quantum devices with statically coupled qubits, we seek realizations of this $\gencx'$ based on CR.

It is known that CR between any two transmons has a preferred direction, in the sense that driving one qubit with the other's resonant frequency effects an operation closer to the ideal CR unitary in equation~\eqref{eqn:ideal_cr_unitary} than doing the reverse. Typical reasons for this asymmetry include qubit frequency collisions and the existence of undesired two-level systems in the device. Even without such error sources, the magnitudes of the $\psi_l$ parameters in equation~\eqref{eqn:ideal_cr_unitary} depend on the signed difference of the resonance frequencies of the control and target~\cite{Malekakhlagh2020-fc}, which means that Rabi oscillation of the target is generated more effectively in the CR drive in one direction than in the other. We thus developed two distinct gate sequences (``forward'' and ``backward'') for $\gencx'$, where the driven system is \qctwo in the former and \qt in the latter. The Toffoli gate can then be implemented with high fidelity on different combinations of transmons.

It is worth mentioning that for CX gates between two qubits, the ``backward'' gate sequence is obtained by simply interposing the ``forward'' CX gate from the target qubit to the control qubit between pairs of Hadamard gates on both qubits. However, such simple decomposition is not viable when qutrits are involved.

\subsection{Forward generalized CX}\label{subsec:forward_gencx}

The forward $\gencx'$ gate, with the CR drive on the qutrit, uses one of the CyCR-like sequences $M_{k}$ in Figure~\ref{fig:gencx_ccr}. Unlike the original CyCR in Figure~\ref{fig:u_ccr}, the repeated units in these sequences are ECR gates $\Wcr{k}^{\pm}$ with matching drive polarity, shown in the bottom row of the same figure.

The use of $\Wcr{k}^{\pm}$ gates guarantees that condition \eqref{eqn:ccr_cx_condition} is satisfiable regardless of the values of $\psi_l$ found in the CR unitary in equation~\eqref{eqn:ideal_cr_unitary}. For example, adapting the expression for $\Usub{CR}^{\pm}$ in equation~\eqref{eqn:ideal_cr_unitary} to a qutrit-qubit system,
\begin{equation}
\label{eqn:wcr}
\begin{split}
  \Wcr{0}^{\pm} = & \X{0} \Usub{CR}^{\pm} \X{0} \Usub{CR}^{\pm} \\
                = & (\proj{0} + \proj{1}) \otimes e^{-\half[i] [(\phi_0 + \phi_1) I \pm (\psi_0 + \psi_1) X]} \\
                & + \proj{2} \otimes e^{-i [\phi_2 I \pm \psi_2 X]},
\end{split}
\end{equation}
and therefore
\begin{equation}
\label{eqn:ccr0}
\begin{split}
  M_{0} = & \Xplus \Wcr{0}^{+} \Xplus \Wcr{0}^{-} \Xplus \Wcr{0}^{+} \\
                    = & (\proj{0} + \proj{2}) \otimes e^{-i \psi_2 X} \\
                    & + \proj{1} \otimes e^{-i (\psi_0 + \psi_1 - \psi_2) X},
\end{split}
\end{equation}
where a global phase was omitted from the last equation. In general, $\psi_0 + \psi_1 - \psi_2 \neq \psi_2$, and all $\psi_l$s depend on the CR pulse duration approximately linearly. Thus, by adjusting the pulse duration, it is possible to create a difference of $\pm\pi$ between the $X$ rotation angles of the first and second terms of equation~\eqref{eqn:ccr0}. The offset rotations of $\theta := 2\psi_2$ can then be cancelled by the $R_x$ gate in Figure~\ref{fig:gencx_ccr}.

In practice, for a given combination of transmons, whichever of $M_0$ and $M_1$ that requires a shorter pulse duration to effect the angle difference of $\pm\pi$ is used for the $\gencx'$ gate. The sign of the phase in $P_1$ and the $\theta$ angle in Figure~\ref{fig:gencx_ccr} follow this choice.

The analysis above is based on the small-$\Delta_l$ approximation made for equation~\eqref{eqn:ideal_cr_unitary}, which is not always justifiable. However, since $\Wcr{}$ is an ECR sequence, effects on it from $\Delta_l$ can in fact be suppressed via rotary pulses, such that equation~\eqref{eqn:wcr} and consequently \eqref{eqn:ccr0} actually hold well.

A forward $\gencx'$ thus requires calibrations of the duration of the CR pulses, the amplitude of the rotary pulse, and the angle of the offset-canceling $R_x$ gate. Details on how the calibrations are performed can be found in Appendix~\ref{app:calibration}.

Incidentally, a sequence of three equispaced $\Xplus$ gates with no interleaved $\ket{1}$-$\ket{2}$ space gates is affected by qutrit detuning by an overall $\Pop{2}{-3\delta \tau}$, where $\tau$ is the interval between the $\Xplus$ gates. Therefore, the $M_k$ sequences are automatically compatible with the form of equation~\eqref{eqn:gencx_prime}, making this a valid implementation of $\gencx'$.

\subsection{Backward generalized CX}

\begin{figure}
  \begin{subcaptiongroup}
    (a)
    \phantomcaption\label{fig:gencx_ccr}
    \begin{minipage}[t]{0.92\linewidth}
    \begin{quantikz}[column sep=0.2cm]
      \lstick{\qctwo} & \gate[2]{M_{0/1}} & \gate{\Pop{1}{\mp \half[\pi]}} & \qw \\
      \lstick{\qt}   &                         & \gate{\Rop{x}{-\theta}}        & \qw
    \end{quantikz}
    \end{minipage}

    \hfill\vspace{4pt}\\

    \begin{minipage}[t]{\linewidth}
    \begin{footnotesize}
    \begin{quantikz}[column sep=0.2cm]                    
     & \gate[2]{M_{0/1}} & \qw \\
     &                         & \qw
    \end{quantikz}
    =
    \begin{quantikz}[column sep=0.2cm]
     & \gate[2]{\Wcr{0/1}^{+}} & \gate{\Xplus} & \gate[2]{\Wcr{0/1}^{\mp}} & \gate{\Xplus} & \gate[2]{\Wcr{0/1}^{\pm}} & \gate{\Xplus} & \qw \\
     &                           &               &                         &               &                           &               & \qw
    \end{quantikz}
    \end{footnotesize}
    \end{minipage}

    \hfill\vspace{4pt}\\

    \begin{minipage}[t]{\linewidth}
    \begin{footnotesize}
    \begin{quantikz}[column sep=0.2cm]
     & \gate[2]{\Wcr{k}^{\pm}} & \\
     &                         &
    \end{quantikz}
    =
    \begin{quantikz}[column sep=0.2cm]
     & \gate[2]{\Usub{CR}^{\pm}} & \gate{\X{k}} & \gate[2]{\Usub{CR}^{\pm}} & \gate{\X{k}} & \\
     &                           &              &                           &              &
    \end{quantikz}
    \end{footnotesize}
    \end{minipage}
  \end{subcaptiongroup}

  \hfill\vspace{8pt}\\

  \begin{subcaptiongroup}
    (b)
    \phantomcaption\label{fig:gencx_rzxpi}
    \begin{minipage}[t]{0.92\linewidth}
    \begin{footnotesize}
    \begin{quantikz}[column sep=0.2cm]
      \lstick{\qctwo} & \gate{\Xplus} & \gate[2]{\Rop{xz}{\half[\pi]}} & \gate[2]{\Rop{xz}{\half[\pi]}} & \gate{\Rop{x}{-\pi}} & \gate{\Xminus} & \qw \\
      \lstick{\qt}   & \gate{H}      &                                &                                & \gate{\Rop{z}{\pi}}  & \gate{H}       & \qw
    \end{quantikz}
    \end{footnotesize}
    \end{minipage}
    \hfill \\
    \begin{minipage}[t]{\linewidth}
    \begin{footnotesize}
    =
    \begin{quantikz}[column sep=0.2cm]
     & \gate{\Xplus} & \gate[2]{\Rop{xz}{\half[\pi]}} & \gate[2]{\Rop{xz}{\half[\pi]}} & \gate{\Rop{x}{-\pi}} & \gate{\Xplus} & \wr\wr & \gate{\Xplus} & \qw \\
     & \gate{H}      &                                &                                & \gate{\Rop{z}{\pi}}  &               & \wr\wr & \gate{H}      & \qw
    \end{quantikz}
    \end{footnotesize}
    \end{minipage}
  \end{subcaptiongroup}
  \caption{Generalized CX gate implementations. (a) Forward-direction $\gencx'$ gate (top) and its components (lower two lines). (b) Backward-direction $\gencx'$ gate and its basis-cycled form.}
\end{figure}
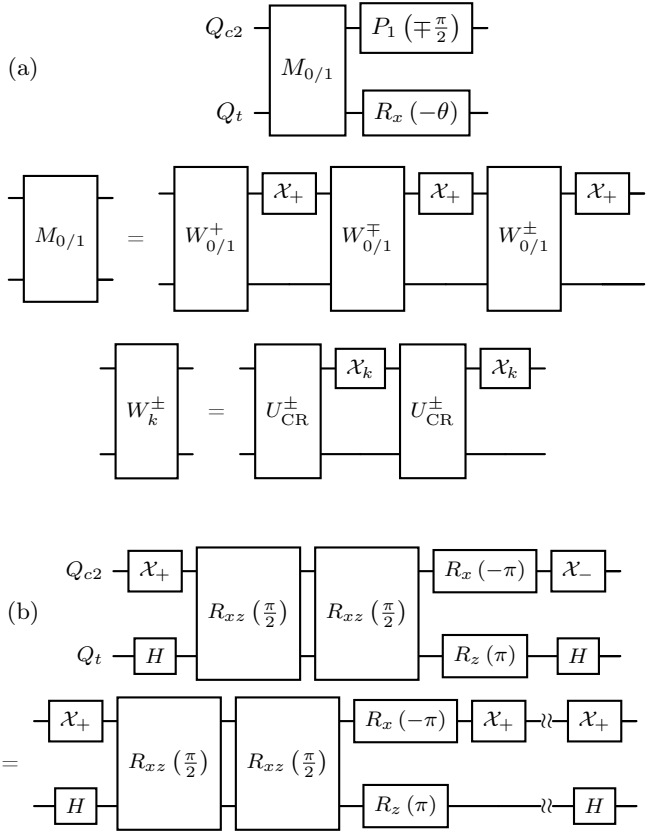

In contrast to the forward $\gencx'$ gate, the backward counterpart, shown in Figure~\ref{fig:gencx_rzxpi}, has essentially no new parameters to be calibrated. In the figure, $\Rop{xz}{\half[\pi]}$ represents the core of the standard CX gate acting on the target qubit and the control qutrit, and corresponds to the unitary
\begin{equation}
  \Rop{xz}{\half[\pi]} = \expfn{-\quarter[i\pi] X_0} \otimes \proj{0} \, + \, \expfn{\quarter[i\pi] X_0} \otimes \proj{1},
\end{equation}
with the qutrit operators on the left and qubit operators on the right.
Two applications of $\Rop{xz}{\half[\pi]}$ together with $\Rop{x}{-\pi}$ on the qutrit results in
\begin{equation}\label{eqn:rxzpi_rxminuspi}
\begin{split}
  & \left[\Rop{x}{-\pi} \otimes I\right] \left[\Rop{xz}{\half[\pi]}\right]^2 \\
  = & I \otimes \proj{0} - \left(\proj{0} + \proj{1} - \proj{2}\right) \otimes \proj{1}.
\end{split}
\end{equation}
The negative sign on $\proj{2}$ in the second term of equation~\eqref{eqn:rxzpi_rxminuspi} is the geometric phase acquired by the qutrit by making a full turn around the Bloch sphere of the $\ket{0}$-$\ket{1}$ space. This sign can be shifted onto qutrit $\proj{1}$ using the $\Xpm$ gates and be exploited to realize, up to a global phase,
\begin{equation}
\begin{split}
  & \left[\Xminus \otimes \Rop{z}{\pi} \right] \left[\Rop{x}{-\pi} \otimes I \right] \left[\Rop{xz}{\half[\pi]}\right]^2 \left[\Xplus \otimes I \right] \\
  = & I \otimes I - 2 \proj{1} \otimes \proj{1},
\end{split}
\end{equation}
which is the $\ket{1}$-controlled generalized CZ gate. Placing this gate sequence in between two Hadamard gates on the target qubit completes the generalized CX gate. Note that with the backward interaction, it is actually more natural to implement the CCZ gate than the Toffoli gate.

This form of $R_{xz}$-based generalized CX is however prone to qutrit detuning errors, and is therefore converted to the basis-cycled version of Figure~\ref{fig:gencx_rzxpi} bottom, where the $\Xplus$ gates are separated by equal intervals. The result is an equispaced sequence of three $\Xplus$ gates. As with the $M_k$ sequences of the forward $\gencx'$, the detuning error acts on this sequence as an overall $P_2$.

The only calibration requirement that may not be obvious from the circuit diagram of the backward $\gencx'$ is the correction for phase accumulation in the $\ket{1}$-$\ket{2}$ subspace of the qutrit \qctwo. This relative phase exists because of the rotary drive in the ECR sequence of the $\Rop{xz}{\half[\pi]}$ gates, which is resonant with the $\ket{0}$-$\ket{1}$ transition of \qctwo and therefore causes an AC Stark shift of the $\ket{2}$ level. It should be emphasized that this correction is however not specific to the current generalized CX gate implementation, and is generally required when operating a transmon as a qutrit in a device that uses ECR as one of the basis gates. In fact, corrections for the same effect are applied to the two CX gates between \qcone and \qctwo in the Toffoli gate decomposition.

\subsection{Validation with cloud quantum computers}
\label{subsec:toffoli_results}

Here, we report the results of implementing the gate decomposition based on the backward $\gencx$ on cloud quantum computers provided by IBM. The names and the relevant details of the used devices are given in Table~\ref{tab:devices}. All quantum processors are of the Eagle family, which is the latest architecture featuring statically coupled qubits. Since our $\gencx$ gates are based on CR, they are not directly usable on processors of the newer Heron family with tunable couplers.

Because the Toffoli gate with the backward $\gencx$ is constructed by simply adding Hadamard gates to the CCZ gate, we decided to perform the demonstrations with the CCZ gate. Although the presented results are for specific sets of transmons, we emphasize that the gate decomposition developed in this paper is applicable to any system with CR-based entangling gates. The selected transmons in Table~\ref{tab:devices} have lower readout and entangling gate error rates compared to the other units on each device, according to the calibration data published at the time of demonstration.

\begin{table*}[htbp]
\centering
\caption{List of devices and qubits used for the demonstration of the ternary CCZ gate decomposition. Error rates and coherence times represent the values published at the time of CCZ gate calibration. ECR error is the error rate of the echoed cross resonance sequence within the standard CX gate decomposition. The last two columns represent the gate time of the Toffoli gate under the eight-CX qubit-only decomposition and the CCZ gate using our decomposition (which equals the Toffoli gate time since the Hadamard gates are applied in parallel to other gates in CCZ and incur no additional time). All times are in units of \us.}
\label{tab:devices}
\begin{tabular}{l c c c c c c c c c c c}
\hline \hline
\multirow{2}{6em}{Device} & \multirow{2}{6em}{\qcone, \qctwo, \qt} & \multicolumn{2}{c}{ECR error ($10^{-3}$)} & \multicolumn{3}{c}{$T_1$} & \multicolumn{3}{c}{$T_2$}  & \multirow{2}{4em}{$\sub{T}{CCX, qubit}$} & \multirow{2}{4em}{$\sub{T}{CCZ, qutrit}$} \\
\cline{3-10}
& & \qcone-\qctwo & \qt-\qctwo & \qcone & \qctwo & \qt & \qcone & \qctwo & \qt & \\
\hline
\ibmdev{cusco} & 115, 114, 113 & 6.5 & 5.3 & 168.5 & 213.0 & 205.0 & 44.6 & 112.7 & 338.9 & 4.08 & 3.29 \\
\ibmdev{kawasaki} & 116, 117, 118 & 4.7 & 4.7 & 210.2 & 188.9 & 172.4 & 291.3 & 74.2 & 170.5 & 5.17 & 4.16 \\
\ibmdev{kyiv} & 75, 90, 94 & 5.9 & 5.5 & 243.2 & 186.3 & 245.0 & 162.9 & 204.6 & 194.5 & 4.94 & 3.97 \\
\ibmdev{kyoto} & 16, 26, 27 & 4.9 & 5.4 & 234.3 & 255.1 & 290.8 & 74.0 & 338.2 & 185.4 & 5.82 & 4.68 \\
\hline \hline
\end{tabular}
\end{table*}

On top of the CCZ circuit introduced in the previous section, pairs of X gates are inserted to idle periods of \qcone and \qt for dynamical decoupling, cancelling undesired accumulation of phases on these qubits due to the static qutrit-qubit longitudinal interactions. The static phase shift from the same source experienced by \qctwo is corrected by inserting two $R_{z}$ gates, one after the first CX gate and the other at the end of the full CCZ gate sequence (See Appendix~\ref{app:calibration} for details).
%We however emphasize that these static phase errors are mostly dependent on the strengths of the couplings between \qctwo and the neighboring qubits, and are unaffected by time-shifting qutrit detuning.

The overall gate time, which is dominated by the four ECR sequences and the idle time inserted for qutrit refocusing, is given in the last column of Table~\ref{tab:devices}. The idle time duration roughly corresponds by construction to that of two ECR sequences. The gate time of the eight-CX qubit-only decomposition of the Toffoli gate on the same combinations of qubits are also shown in Table~\ref{tab:devices} for reference.

The fidelity of the gate decomposition is determined via QPT, which is performed by measuring all possible Pauli expectation values of states obtained by applying the CCZ gate to specific initial states. The initial states are given by preparing each of \qcone, \qctwo, \qt in one of the four states $\ket{Z+}$, $\ket{Z-}$, $\ket{X+}$, and $\ket{Y+}$, where $\ket{P\pm}$ is the eigenstate of the local Pauli operator $P$ with eigenvalue $\pm 1$. The total number of measured expectation values are therefore $4^3 \times 3^3 = 1728$, and each is obtained by executing the corresponding circuit for 2000 times. To account for qubit measurement errors, confusion matrix for readout assignment of each qubit is obtained from independent measurements, and the results are used to reconstruct the positive operator-valued measure elements corresponding to each noisy Pauli measurement. The quantum channel $\mathcal{E}$ of the gate sequence is reconstructed from these expectation values, using CVXPY splitting conic solver as the fitter~\cite{diamond2016cvxpy,agrawal2018rewriting}. Process fidelity $\sub{F}{pro}(\mathcal{E}, U)$ between $\mathcal{E}$ and a unitary channel $U$ is defined as
\begin{equation}
  \sub{F}{pro}(\mathcal{E}, U) = \frac{\mathrm{Tr}\left[S_U^{\dagger} S_{\mathcal{E}}\right]}{d^2},
\end{equation}
where $S_{\mathcal{F}}$ is the superoperator representation of channel $\mathcal{F}$ and $d$ is the dimension of the Hilbert space ($d=8$ for three qubits).

The gate fidelity obtained immediately after the calibration of the $\X{1}$ gate and the local $R_z$ corrections on the \ibmdev{kyoto} device is 92.5$\pm$0.1\%, where the uncertainty is statistical. Corresponding values for \ibmdev{cusco} and \ibmdev{kyiv} are 89.1$\pm$0.1\% and 86.6$\pm$0.1\%, respectively. On \ibmdev{kawasaki}, we performed the first fidelity measurement nine hours after the calibration and obtained the gate fidelity of 92.8$\pm$0.1\%.

\begin{figure}
  \centering
  \includegraphics[width=\linewidth]{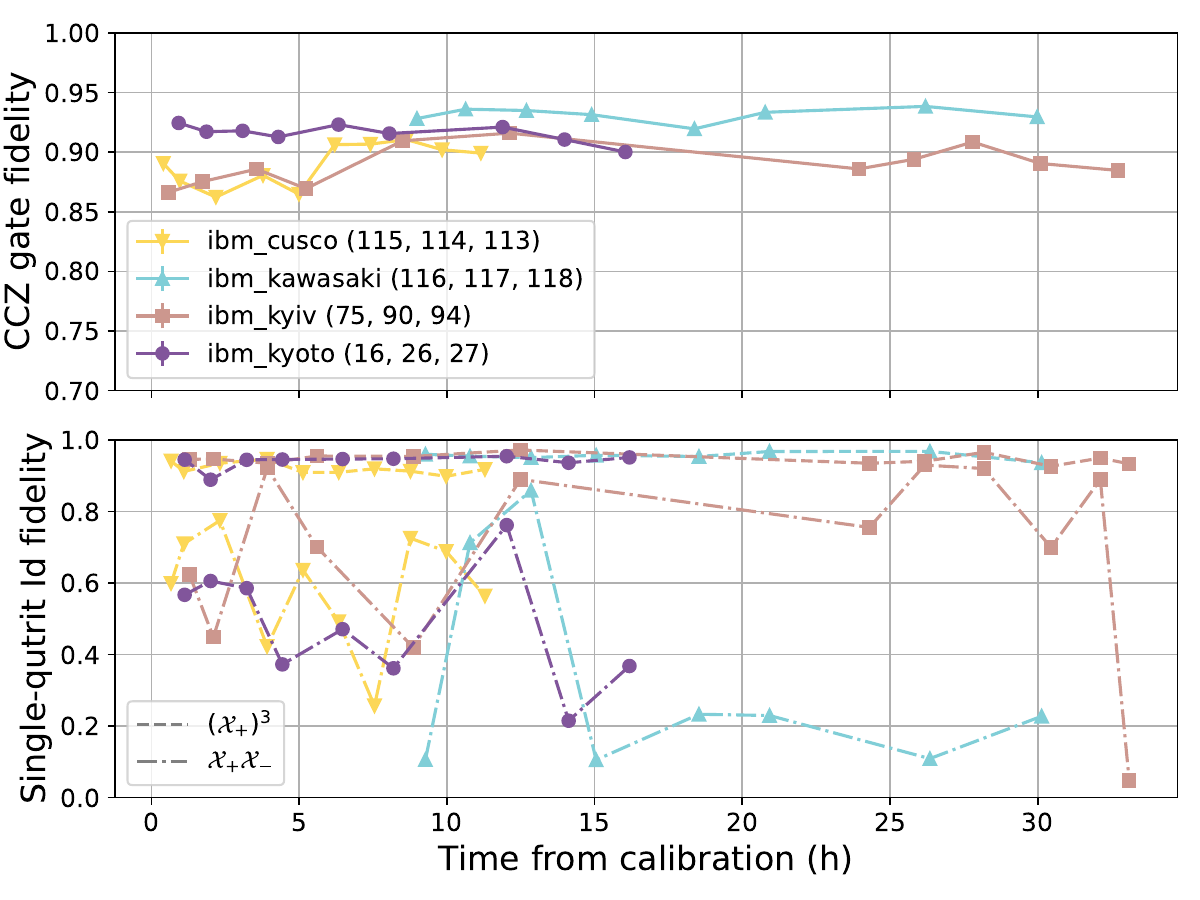}
  \caption{Time variation of the CCZ and the reference identity gate fidelities. The fidelity values of the CCZ gate and $(\Xplus)^3$ sequence remain close to the respective initial values, while those of the $\Xplus\Xminus$ sequence fluctuate widely. Note the difference in the range of the vertical axes of the two graphs.}
  \label{fig:ccz_fidelity}
\end{figure}

To confirm that the gate decomposition is protected against fluctuation of the $\ket{1}$-$\ket{2}$ frequency by basis cycling, the same tomography experiment is repeated multiple times over a period of 33 hours for each device without any recalibration. As shown in the top half of Figure~\ref{fig:ccz_fidelity}, the gate fidelity values remain within a few percent of the respective initial values. In fact, on \ibmdev{kawasaki}, the highest gate fidelity at 93.8$\pm$0.1\% is obtained 26 hours after the calibration. That the underlying accuracy of the $\X{1}$ gate does however fluctuate strongly is verified in the bottom half of the same figure, where gate fidelities of the decompositions of identity into $(\Xplus)^3$ and $\Xplus\Xminus$ are compared. In both decompositions, intervals are inserted between the gates to make the overall duration of the sequences identical to that of the CCZ gate. The $\Xplus\Xminus$ sequence, which is the simplest instance of a transient ternary circuit of Figure~\ref{fig:u_ternary_circuit} top, is not protected against qutrit detuning and indeed exhibits fluctuating gate fidelity values.

As a reference, we measured the gate fidelity of the eight-CX implementation of the Toffoli gate via the same QPT procedure on various qubit combinations of multiple devices, including one with a Heron processor. See Appendix~\ref{app:qubit_toffoli} for the full list of performed measurements. For Eagle processors, the observed fidelity values are in the range of 0.57-0.90, and thus our ternary decomposition of the CCZ gate generally outperforms the qubit-only implementation of the Toffoli gate. On the other hand, fidelity values as high as 0.97 were obtained for the Heron processor, which is a vivid illustration of the improvement in the error rate of the two-qubit gates enabled by tunable couplers. It will be interesting to explore how high the fidelity of a ternary decomposition of the CCZ gate on tunable-coupler devices can be, especially given the favorable scaling in the number of entangling gates of qutrit-based multi-controlled X gates. Such an experiment will however require a development of a generalized CZ gate not based on CR, along the lines of works such as Refs.~\cite{Chu2022-go,Goss2022-kj,Nguyen2024-eo}, that is compatible with basis cycling, which is left for future work.

We then investigate the sources of CCZ gate infidelity through measurements of two sets of quantities. The first set is
\begin{equation}
  P_j = \mathrm{Tr}\left[\proj{j} \mathcal{E}(\proj{j})\right] \quad (j=000, \dots, 111),
\end{equation}
where $\mathcal{E}$ is the quantum channel of the CCZ circuit. These values represent the diagonal entries of the so-called truth table. The second set is
\begin{equation}
  \Phi_{j,k} = \mathrm{arg}\,\mathrm{Tr}\left[\ketbra{k}{j} \mathcal{E}(\ketbra{j}{k})\right] \quad (j,k=000,\dots,111).
\end{equation}
When $\mathcal{E}$ is a diagonal unitary channel as
\begin{equation}
  \mathcal{E}(\rho) = \sum_{lm} e^{i(\phi_l - \phi_m)} \proj{l} \rho \proj{m},
\end{equation}
we have
\begin{equation}
  \Phi_{j,k} = \phi_j - \phi_k,
\end{equation}
i.e., $\Phi_{j,k}$ corresponds to the relative phase between its $j$th and $k$th diagonals of the unitary. These two sets of quantities provide complementary classical fidelities, and therefore their combined measurement corresponds to a low-overhead alternative to performing full process tomography in evaluating the quality of quantum operations~\cite{Hofmann2005-lp}. Measurements of $P_j$ and $\Phi_{j,k}$ are described in Appendix~\ref{app:error_analysis}.

Figure~\ref{fig:error_analysis} (a) shows $P_j$ and $\Phi_j = \Phi_{j,000}$ obtained for transmons 115, 114, and 113 of \ibmdev{cusco}. For an ideal CCZ gate, we expect $P_j = 1$ for all $j$, $\Phi_j = 0$ for $j=001,\dots,110$, and $\Phi_{111} = \pi$. It is revealed that the measured $\Phi_j$ are well aligned with the ideal values, but $P_j$ are between 0.90 and 0.94 for all $j$. This result, which is in fact similar in all tested devices, indicates that the population migration to off-diagonal elements is the major bottleneck for achieving a higher gate fidelity.

Population migration can happen because of qubit and qutrit decoherence and gate errors. We analyze how different parts of the circuit contribute to this effect by measuring $\sub{F}{TT} = \sum_j P_j / d$ $(d=8)$ of different identity circuits derived from the CCZ gate. For diagonal gates like identity and CCZ, this quantity $\sub{F}{TT}$ is equivalent to what is sometimes called the truth-table fidelity.

The first such circuit ($(\Xplus)^6$) consists of six contiguous $\Xplus$ gates applied on \qctwo and dynamical decoupling sequences on \qcone and \qt. The truth-table fidelity of this circuit represents the population migration from errors in the $\X{0}$ and $\X{1}$ gates. The three next circuits, labeled $\mathrm{Id}_{A}$, $\mathrm{Id}_{B}$, and $\mathrm{Id}_{C}$, are different modifications of the CCZ gate, all featuring the same timings of the $\Xplus$ gates. Circuit $\mathrm{Id}_A$ replaces the CX and CZ gates within the CCZ gate with dynamical decoupling sequences. Circuit $\mathrm{Id}_B$ retains the CZ gate but have the two CX gates replaced, while $\mathrm{Id}_C$ retains the two CX gates and replaces the CZ gate. The decline in $\sub{F}{TT}$ going from $(\Xplus)^6$ to $\mathrm{Id}_A$ thus corresponds to the population migration from decoherence, and those between the $\mathrm{Id}_{A,B,C}$ and CCZ circuits arise from errors from the entangling gates. Overall, the results suggest that decoherence is the dominant source of infidelity, with non-negligible contributions from errors in the single-qutrit and entangling gates.

\begin{figure}
  \centering
  \includegraphics[width=\linewidth]{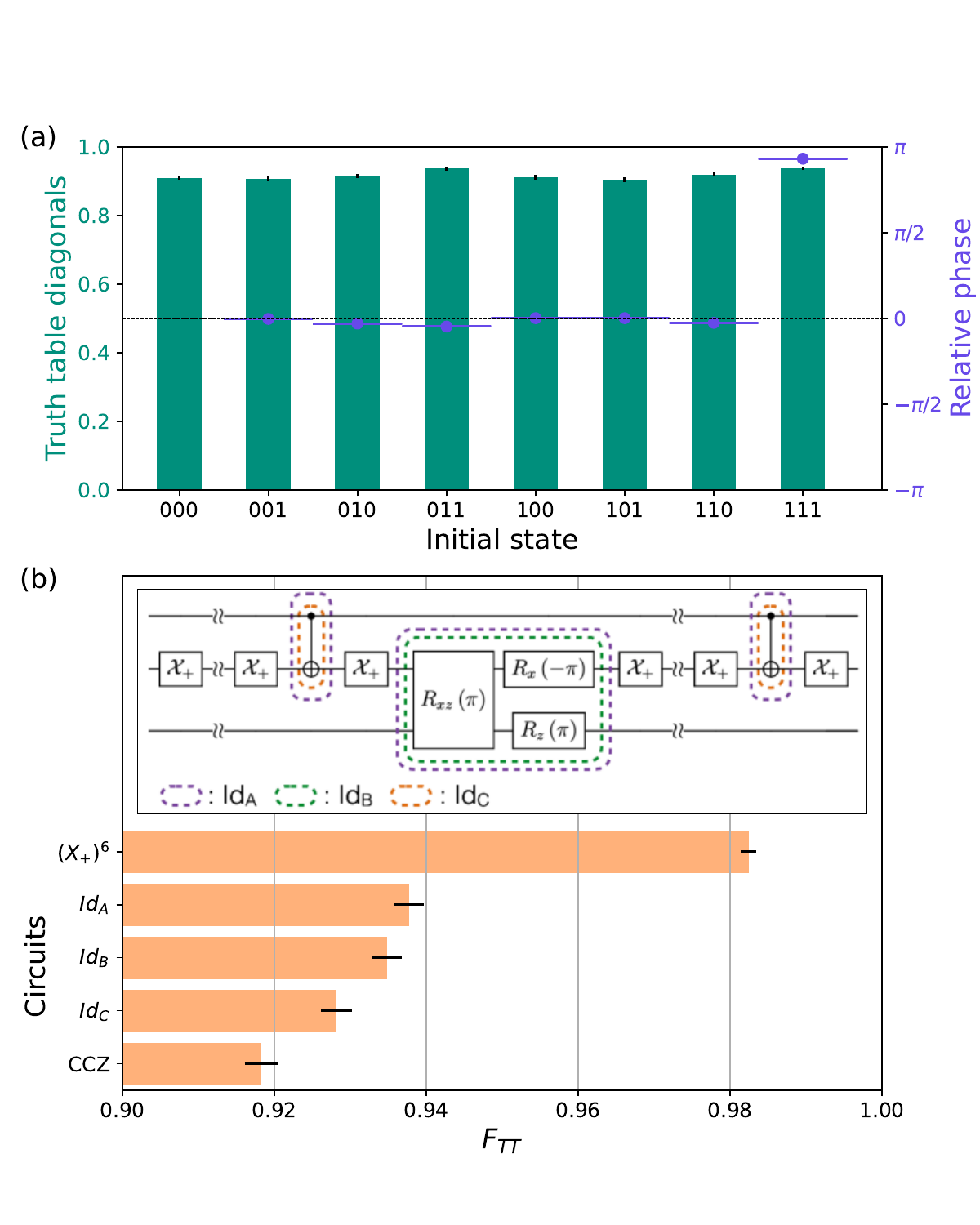}
  \caption{(a) Diagonal entries of the truth table $P_j$ (bar, left scale) and element-wise relative phases $\Phi_j$ (line, right scale) of the CCZ gate. The states are labeled in the order of \qt,\qctwo,\qcone. Error bars represent statistical uncertainties, and are smaller than the markers for the relative phases. (b) Truth-table fidelity $\sub{F}{TT}$ of various identity decompositions and the CCZ circuit. The inset diagram shows which components of the CCZ circuit are replaced by identity sequences for the three $\mathrm{Id}$ circuits.}
  \label{fig:error_analysis}
\end{figure}

%% file: discussion.tex
In this paper, we formulated a generalization of the qubit echo protocol to cancel undesired phase accumulation on general qudit systems. The proposed technique, which we call basis cycling, requires only the $\pi$-pulse gates for all neighboring two-level subspaces of the qudit, and works by cyclically permuting the qudit computational basis states. Its operational principle is straightforward, allowing solutions to various problems to be designed effortlessly. It is also made clear that the application of the technique is in fact not limited to phase cancellation, and in some cases interesting unitary operations on the system coupled to the qudit can be engineered through basis cycling.

We applied these findings to a ternary decomposition, i.e., one involving a qutrit, of the Toffoli and CCZ gates on a statically coupled fixed-frequency superconducting quantum computer. The original gate decomposition using two single-qutrit gates is modified to a basis-cycled sequence of three single-qutrit cyclic gates to dynamically cancel the phase errors arising from qutrit detuning. Additionally, we devised two generalized CX gate implementations based on cross resonance interactions embedded within basis cycling sequences. Using these components, we achieved a gate fidelity of 93.8$\pm$0.1\%, measured through quantum process tomography, for the CCZ gate on an IBM quantum computer. As a reference, the gate fidelity of the most efficient known qubit-based Toffoli gate decomposition was measured on multiple qubit combinations in various IBM quantum devices. Our CCZ fidelity value was higher than that obtained for any of the qubit combinations in devices whose entangling gates are based on cross resonance. Furthermore, thanks to basis cycling, the gate fidelity was shown to remain high for at least 33 hours without recalibration of the qutrit control gates, even though the resonant frequency of the qutrit fluctuates at a much shorter time scale.

Our calibration program for the qutrit gates and implementations of the Toffoli and CCZ gates are general and in principle usable on any qubit combination on any device supporting pulse-level control via Qiskit. To underline the generality and versatility of basis cycling, which are the central claims of this paper, we have made the entire code base public, and invite the readers to contrive novel applications of qutrits using this framework.

%% file: acknowledgments.tex
We thank Will Shanks for crucial support enabling the experiments on IBM devices and Shogo Toma for providing theoretical insights. YI is supported in part by Japan Society for the Promotion of Science (JSPS), through Grants-in-Aid for Scientific Research (KAKENHI) Grant No. 20K22347. This work is supported by IBM-UTokyo lab. IBM, the IBM logo, and ibm.com are trademarks of International Business Machines Corp., registered in many jurisdictions worldwide. Other product and service names might be trademarks of IBM or other companies. The current list of IBM trademarks is available at https://www.ibm.com/legal/copytrade.

%% file: qudit_phase.tex
Modeling a transmon as an weakly anharmonic oscillator circuit with up to $d$ energy levels that is capacitively coupled to a classical oscillating voltage source, we can write down its Hamiltonian as
\begin{equation}
  \sub{H}{tr} = H_0 + \sub{H}{d},
\end{equation}
with
\begin{equation}
  H_0 = \sum_{l=0}^{d-1} \hbar \Omega_l \proj{l}
\end{equation}
and
\begin{multline}
  \sub{H}{d} = i \sub{A}{d}(t) \sin (\sub{\omega}{d} t + \sub{\phi}{d}) \\
  \sum_{l=0}^{d-2} \alpha_l \left(\ketbra{l}{l + 1} - \ketbra{l + 1}{l}\right).
\end{multline}
In the expressions, $\Omega_l$ and $\ket{l}$ are the $l$th energy eigenvalue and eigenstate of the free Hamiltonian $H_0$, $\sub{A}{d}(t)$, $\sub{\omega}{d}$, and $\sub{\phi}{d}$ are the amplitude, angular frequency, and offset phase of the drive voltage, and $\alpha_l \sim \sqrt{l + 1}$ is the transition amplitude for the $l$th level. By moving into the qudit frame and applying the rotating-wave approximation, $\sub{H}{tr}$ transforms to
\begin{equation}\label{eqn:rwa_hamiltonian}
  \bar{H}_{\mathrm{tr}} = \half[\sub{A}{d}(t)] K(\sub{\omega}{d}, \sub{\phi}{d}),
\end{equation}
where
\begin{equation}\label{eqn:htr_drive_operator}
  K(\sub{\omega}{d}, \sub{\phi}{d}) = \sum_{l=0}^{d-2} \alpha_l \left(e^{-i(\omega_l - \sub{\omega}{d}) t} e^{i\sub{\phi}{d}} \ketbra{l}{l + 1} + \mathrm{h. c.} \right),
\end{equation}
with $\omega_l := \Omega_{l+1} - \Omega_{l}$.

With particular attention to the drive phase factors, $K(\sub{\omega}{d}, \sub{\phi}{d})$ can be rewritten in terms of the phase gradation operator
\begin{equation}
  Q(\phi) = \sum_{l=0}^{d-1} e^{il\phi} \proj{l}
\end{equation}
as
\begin{equation}\label{eqn:virtual_q}
  K(\sub{\omega}{d}, \sub{\phi}{d}) = Q(-\sub{\phi}{d}) K(\sub{\omega}{d}, 0) Q(\sub{\phi}{d}).
\end{equation}
For $d=2$, $Q(\phi) = \Rop{z}{\phi}$ up to a global phase, and equation~\eqref{eqn:virtual_q} describes the familiar principle of virtual-Z gate: Shifting the phase of the control drive of a qubit gate by $\phi$ is equivalent to applying an $\Rop{z}{\phi}$ gate immediately before the gate. A similar statement can be made in the general qudit context because
\begin{equation}
   K(\sub{\omega}{l}, 0) \sim
   \begin{cases}
   \alpha_l X_l & \text{if } \sub{\omega}{d} \sim \omega_l \\
   0 & \text{otherwise}
   \end{cases}
\end{equation}
when oscillating terms in the Hamiltonian are ignored, which implies that in equation~\eqref{eqn:virtual_q} only terms for neighboring two levels in $Q(\phi)$ are relevant at any time. The phase of the drive signal therefore always sets the phase difference between the two levels it addresses.

Practically, then, to give a physical implementation of a qudit circuit on superconducting quantum computers, the logical circuit must first be analyzed to track the phase differences in individual two-level systems of each qudit as they are updated via diagonal unitary gates. Each drive pulse effecting transitions between two levels must then be given the phase corresponding to the cumulative phase gap of the levels. For example, in the following circuit
\begin{center}
\begin{quantikz}[column sep=0.2cm]
  & \gate{\Rop{z(k)}{\phi}} & \gate{\X{k}} & \gate{\Rop{z(k+1)}{\chi}} & \gate{\X{k+1}} & \qw
\end{quantikz}
\end{center}
with
\begin{equation}
  \Rop{z(k)}{\phi} = e^{-\half[i \phi]} \proj{k} + e^{\half[i \phi]} \proj{k + 1} + \sum_{l \neq k, k+1} \proj{l},
\end{equation}
the $R_{z(k)}$ gate creates a phase gap of $\phi$ between levels $k$ and $k+1$ and $-\half[\phi]$ between $k+1$ and $k+2$, and the $R_{z(k+1)}$ gate opens the gap between $k+1$ and $k+2$ by an additional $\chi$. The phase gap between $k+2$ and $k+3$ levels created by $R_{z(k+1)}$ has no effect since there is no other gate in the circuit that addresses this two-level system. The control signal for the $\X{k}$ and $\X{k+1}$ gates are therefore given phases $\phi$ and $-\half[\phi] + \chi$, respectively.

We now see that any diagonal single-qudit unitary can be implemented by combining phase shifts on the control drives for different two-level systems. One class of operators that are useful when designing qudit circuits is the $\Pop{l}{\phi}$ ($l=0,1,\dots,d-1$) gates, which apply a phase factor of $e^{i\phi}$ to the $l$th level. The physical implementation of $\Pop{0}{\phi}$ and $\Pop{d-1}{\phi}$ are particularly simple, since they involve shifting the phase of the drive signals for only one two-level system (the lowest and the highest, respectively). A general $\Pop{l}{\phi}$ is implemented as phase shifts of $\phi$ and $-\phi$ of the $\ket{l-1}$-$\ket{l}$ and $\ket{l}$-$\ket{l+1}$ control signals, respectively.

Physical effects such as longitudinal coupling and AC Stark shifts, as well as miscalibration of the drive frequency, cause the drive frames to be detuned from the qudit frame. Small detuning appears as a time-dependent phase offset of the control pulses, as evident from equation~\eqref{eqn:htr_drive_operator} for $\sub{\omega}{d} = \omega_k + \delta$:
\begin{equation}
\begin{split}
  K(\omega_k + \delta, \sub{\phi}{d}) = & \sum_{l=0}^{d-2} \alpha_l \left(e^{-i(\omega_l - \omega_k - \delta) t} e^{i\sub{\phi}{d}} \ketbra{l}{l + 1} + \mathrm{h. c.} \right) \\
  \sim & \alpha_k \left( e^{i(\sub{\phi}{d} + \delta t)} \ketbra{k}{k + 1} + \mathrm{h. c.} \right) \\
  = & Q(-\sub{\phi}{d} - \delta t) X_k Q(\sub{\phi}{d} + \delta t).
\end{split}
\end{equation}

%% file: reframing_echo.tex
The qubit echo can serve multiple purposes depending on the structure of the $\upj{V}$ unitaries in equation~\eqref{eqn:qubit_vj_decomposition}. In the following, we give three examples in detail.

\subsection{Refocusing}

When there is a small detuning $\delta$ between the drive and qubit resonance frames, the echo sequence may be used to refocus the qubit control pulses. First, under such a detuning, the so-called $\pi$ pulse that is supposed to realize the $X$ gate at time $t$ instead effects
\begin{equation}
\begin{split}
  \up{X}{t} & = e^{i\frac{\delta t}{2} Z} X e^{-i\frac{\delta t}{2} Z} \\
  & = X e^{-i\delta t Z}.
\end{split}
\end{equation}
If this phase error accumulates on a quantum circuit such that its overall unitary has a form
\begin{equation}
  \Usub{detuned} = e^{-i \delta \alpha Z} \Usub{ideal}
\end{equation}
with $\Usub{ideal}$ the unitary without the phase error, one can append an echo sequence to the circuit with an interval of $\alpha$ in between to achieve
\begin{equation}\label{eqn:qubit_refocusing}
  \up{X}{t_0+\alpha}\up{X}{t_0} \Usub{detuned} = \Usub{ideal}.
\end{equation}

To see this, we recast the left hand side of equation~\eqref{eqn:qubit_refocusing} as
\begin{multline}
  \up{X}{t_0+\alpha}\up{X}{t_0} \Usub{detuned} \\
  = \left(X e^{-i\delta (t_0 + \alpha) Z} X e^{-i\delta t_0 Z} e^{-i\delta \alpha Z}\right) \Usub{ideal}.
\end{multline}
The parenthesized expression on the right hand side is in the form of equation~\eqref{eqn:qubit_echo_unitaries} and trivially seen to evaluate to identity.

\subsection{Dynamical decoupling}

When a qubit evolves under a unitary $\Usub{diag}$ that is diagonal in the computational space of the qubit, it may be possible to decompose $U$ into a product of two unitaries $\up{U}{0}$ and $\up{U}{1}$. The most obvious instance of such an evolution is when there is no gate directly acting on the qubit for some finite duration but the qubit is subject to some longitudinal coupling with neighboring qubits. Let
\begin{equation}
  \up{U}{j} = e^{i\upj{\phi}_0} \proj{0} \otimes \upj{u}_0 + e^{i\upj{\phi}_1} \proj{1} \otimes \upj{u}_1.
\end{equation}
If we desire to disentangle this qubit from the environment, and if $\up{u}{1}_1 \up{u}{0}_0 = \up{u}{1}_0 \up{u}{0}_1 =: u$ holds exactly or approximately, we can alter the qubit evolution by inserting two X gates as
\begin{equation}
  \begin{split}
  \Usub{DD} & = e^{i(\up{\phi}{1}_1 + \up{\phi}{0}_0)} \proj{0} \otimes \up{u}{1}_1 \up{u}{0}_0 \\
  & \quad + e^{i(\up{\phi}{1}_0 + \up{\phi}{0}_1)} \proj{1} \otimes \up{u}{1}_0 \up{u}{0}_1 \\
  & = \left(e^{i(\up{\phi}{1}_1 + \up{\phi}{0}_0)} \proj{0} + e^{i(\up{\phi}{1}_0 + \up{\phi}{0}_1)} \proj{1}\right) \otimes u.
  \end{split}
\end{equation}
The qubit evolution is indeed decoupled, albeit with nontrivial phase accumulation without further assumptions on the structure of $\upj{U}$.

\subsection{Echoed cross resonance}

When a control qubit is driven at the resonant frequency of the target qubit, the resulting unitary of the control-target system has a structure
\begin{equation}
  \Usub{CR}^{\pm} = \sum_{l=0}^{1} \proj{l} \otimes \expfn{-\half[i] \left[ \phi_l I \pm \psi_l X + \theta_l Z \right]}.
\end{equation}
The $\phi_l$ terms arise from the AC Stark shift of the control qubit resonance. The ECR sequence without the rotary tones is then
\begin{equation}
  \begin{split}
  \Usub{ECR} & = X \Usub{CR}^{-} X \Usub{CR}^{+} \\
  & = e^{-\half[i] (\phi_0 + \phi_1)} \left\{ \proj{0} \otimes e^{-\half[i] \left[ -\psi_1 X + \theta_1 Z\right]} e^{-\half[i] \left[ \psi_0 X + \theta_0 Z\right]} \right. \\
  & \qquad \left. + \proj{1} \otimes e^{-\half[i] \left[ -\psi_0 X + \theta_0 Z\right]} e^{-\half[i] \left[ \psi_1 X + \theta_1 Z\right]} \right\}.
  \end{split}
\end{equation}
Thus the AC Stark phases are factored out as a global phase, and if $\theta_j$ are negligible, the angles of $X$ rotations accompanying $\proj{0}$ and $\proj{1}$ are of equal magnitude and opposite sign. The rotary in fact create an equivalent  situation to $\theta_j \sim 0$, by cancelling the overall effect of the $Z$ terms.

%% file: qudit_hcr.tex
To determine the terms contained in the effective Hamiltonian of cross resonance in the general qudit-qudit system, we follow the analysis in Ref.~\cite{Malekakhlagh2020-fc}, where time-dependent Schrieffer-Wolf perturbation theory is employed to derive the qubit-qubit effective Hamiltonian. The reference in fact first derives the effective Hamiltonian in the general qudit-qudit space before truncating the result to 2$\times$2 dimensions. Without such truncation, we find that for a CR drive resonant with the $\ket{k}$-$\ket{k+1}$ transition of the target qudit, the terms in the effective Hamiltonian are generated from the products of the following two expressions and their time integrals:
\begin{multline}
K_J = J \sum_{l, m} {\sub{\nu}{c}}_l {\sub{\nu}{t}}_m e^{-i ({\sub{\omega}{c}}_l - {\sub{\omega}{t}}_m) t} \ketbra{l}{l+1} \otimes \ketbra{m + 1}{m} \\
   + \mathrm{h.c.},
\end{multline}
\begin{equation}
  K_{\Omega} = \half[\Omega] \sum_{l} {\sub{\nu}{c}}_l e^{-i ({\sub{\omega}{c}}_l - {\sub{\omega}{t}}_k) t} \ketbra{l}{l+1} \otimes I + \mathrm{h.c.}
\end{equation}
The operators on the control qudit appear on the left of the tensor product symbol. The transition amplitudes ${\sub{\nu}{c/t}}_l$ and frequencies ${\sub{\omega}{c/t}}_l$ are for the the transition between $l+1$th and $l$th levels of the control and target qudits, respectively.

Only the static terms appear in the effective Hamiltonian. The lowest-order terms are obtained from $K_J^2$, $K_{\Omega}^2$, and $K_J K_{\Omega}$. Most of these terms are diagonal in both the control and target systems. The only non-diagonal elements arise from $K_J K_{\Omega}$ by combining terms proportional to $e^{\pm i {\sub{\omega}{t}}_k t}$, which results in a product of a diagonal operator on the control qudit and an $X_k$ on the target qudit. Therefore, if the target space is truncated at two levels, we obtain an effective Hamiltonian of equation~\eqref{eqn:qudit_hcr}.

%% file: calibration.tex
\subsection{\texorpdfstring{$\X{1}$}{X1} gate}

 The $\X{1}$ gate is implemented as a Gaussian DRAG pulse~\cite{Motzoi2009-oi,Gambetta2011-gr} with the duration and the pulse width parameter aligned with the $\X{0}$ gate pulse precalibrated on the device. The parameters to be calibrated are therefore the $\ket{1}$-$\ket{2}$ resonance frequency $f_{12}$ and the amplitude and the $\beta$ parameter of the Gaussian DRAG pulse. The calibration sequence is similar to a typical qubit calibration, following a two-iteration procedure for each of the parameters. In general, the first iteration involves a parameter scan to find a rough estimate, and the second iteration performs an experiment designed to measure the small difference between the true parameter value and this estimate.

\begin{figure}
  \centering
  \begin{quantikz}[column sep=0.15cm]
   \lstick{$\ket{0}$} & \gate{\X{0}} & \gate{\SX{1}} & \gate{\mathrm{Delay}(\tau)} & \gate{\Rop{z(1)}{\phi}} & \gate{\SX{1}} & \gate{\X{0}} & \meter{} \\
  \end{quantikz}
  \caption{Circuit diagram for $f_{12}$ measurement. Delay $\tau$ is fixed but the angle $\phi$ is varied in small steps. Probability of observing 1 as a function of $\phi$ is then fit with a cosine function, whose phase offset corresponds to $2\pi \Delta f_{12} \tau$, where $\Delta f_{12}$ is the average detuning between the true $f_{12}$ and the drive frame.}
  \label{fig:fine_frequency_circuit}
\end{figure}
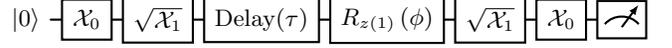

 One difference with respect to the qubit calibration sequence is in the second iteration of the $f_{12}$ determination. In the case of a qubit, one would perform a Ramsey experiment, whereby the qubit is left on the equator of the Bloch sphere for a varying duration, and the frequency offset between the qubit and the drive is inferred from the phase drift frequency. However, a similar experiment in the $\ket{1}$-$\ket{2}$ space would not produce an accurate result due to the low-frequency fluctuation of the true $f_{12}$ itself. This fluctuation causes the observed Ramsey oscillation pattern to have multiple Fourier components, with the pattern often corrupted for longer delay durations due to qutrit decoherence. We therefore instead measure the phase offset at a fixed duration by an alternative Ramsey experiment using circuits depicted in Figure~\ref{fig:fine_frequency_circuit}. The frequency value obtained from this method corresponds to the mean of the fluctuating $f_{12}$.

After the $\X{1}$ pulse is calibrated to produce a $\pi$ rotation that is as pure as possible in the $\ket{1}$-$\ket{2}$ space, AC Stark-induced phase offsets on the $\ket{0}$ level due to the $\X{1}$ pulse and on the $\ket{2}$ level due to the $\X{0}$ and $\SX{0}$ pulses are measured. The $P_0$ and $P_2$ gates to correct these phase offsets are appended to all occurrences of $\X{0}$ and $\X{1}$ gates on the qutrit.

While the Toffoli pulse sequence only utilizes $\X{1}$, the calibration experiments require an implementation of $\SX{1}$, which is therefore calibrated as a byproduct. Using $\X{1}$ and $\SX{1}$ together with $\X{0}$, $\SX{0}$, $R_{z(0)}$, and $R_{z(1)}$ gates, we can perform a qutrit interleaved randomized benchmarking~\cite{Morvan2021-iy} of the $\X{1}$ gate. The obtained error-per-Clifford value is as low as $(2.2 \pm 0.5) \times 10^{-4}$ on some qutrits.

\subsection{CyCR for forward generalized CX}

For the $M_k$ gate in Section~\ref{subsec:forward_gencx} to function as a CyCR sequence, the $\Wcr{k}^{\pm}$ gates must have the form
\begin{equation}\label{eqn:desired_wcr}
  \Wcr{k}^{\pm} = \sum_{l=0}^{2} \proj{l} \otimes \expfn{-\half[i] (\phi_l I \pm \psi_l X)}.
\end{equation}
This in turn requires careful determination of the parameters of the CR pulse and the simultaneous counter pulse on \qctwo and \qt, respectively.

To a good approximation, a general CR pulse with a sufficiently low amplitude effects the following qutrit-diagonal unitary operation on a qutrit-qubit system:
\begin{equation}\label{eqn:physical_ucr}
  \Usub{CR}^{\pm} = \sum_{l=0}^{2} \proj{l} \otimes \expfn{-\half[i] \left[ \phi_l I \pm \psi_l X \pm \lambda_l Y + \chi_l Z \right]}.
\end{equation}
On some transmon combination, at high pulse amplitude, certain nonlinear effects become prominent and breaks down the qutrit-diagonality of the unitary. On the other hand, finite coherence times of the transmons dictate that faster qubit rotations, which generally results from higher drive amplitude, are preferrable. We thus seek the highest amplitude that preserves the diagonality of the unitary.

Equation~\ref{eqn:physical_ucr} features terms with the $Y$ operator, which is absent in equations~\eqref{eqn:qudit_hcr} and \eqref{eqn:ideal_cr_unitary}. In a practical physical implementation of the quantum computer, each drive signal line connecting an output port of room-temperature electronics and a cryogenic qubit has a distinct optical length. This causes the qubit control signals to have different phase offsets at the qubits even if they are synchronized at the output. In particular, the CR drive, applied on the control qubit, is therefore phase-shifted from the drive signal on the target qubit by an a priori unknown value. Furthermore, due to the existence of the $Z$ terms that are dependent on $l$ in equation~\eqref{eqn:qudit_hcr}, the target qubit experiences rotations in the Bloch sphere about slightly different axes during the pulse ramp-up and ramp-down of the CR drive depending on the control qutrit state. As a result, $\lambda_l$ in equation~\eqref{eqn:physical_ucr} has a minor depedence on $l$.

Once a CR drive amplitude is roughly determined, the calibration strategy for $M_0$ is to first set $\lambda_2 = \chi_2 = 0$, then suppress the remaining non-X terms in $\Wcr{0}^{\pm}$ through rotary pulsing. The strategy for $M_2$ is similar and proceeds by eliminating $\lambda_0$ and $\chi_0$. Because pulse durations must be given as a multiple of a finite unit of time on IBM devices, we perform a fine calibration of the CR pulse amplitude at the end so that the Rabi oscillation angles of the qubit are separated by $\pm\pi$ between when the control is in $\ket{0}$ state and when it is in $\ket{1}$ state.

Condition $\lambda_2 = 0$ can be achieved trivially by adjusting the CR drive phase. Then, an application of a off-resonant Stark tone~\cite{Wei2022-us} can be used to create an $l$-independent $Z$ term in its Hamiltonian, with which $\chi_2$ is cancelled. This is a similar technique to Floquet engineering of a single qubit presented in Ref.~\cite{Nguyen2024-eo}. The overall unitary of the combined CR and Stark pulses at this stage is
\begin{multline}
  \Usub{CR}^{\pm'} = \sum_{l=0}^{1} \proj{l} \otimes \expfn{-\half[i] \left[ \phi_l I \pm \psi'_l X \pm \lambda'_l Y + \chi'_l Z \right]} \\
  + \proj{2} \otimes \expfn{-\half[i] \left[\phi_2 I \pm \psi_2 X\right]},
\end{multline}
and therefore the $\Wcr{0}^{\pm}$ unitary is
\begin{multline}
  \Wcr{0}^{\pm} = \proj{0} \otimes \prod_{k=0,1} \expfn{-\half[i] \left[ \phi_k I \pm \psi'_k X \pm \lambda'_k Y + \chi'_k Z \right]} \\
  + \proj{1} \otimes \prod_{k=1,0} \expfn{-\half[i] \left[ \phi_k I \pm \psi'_k X \pm \lambda'_k Y + \chi'_k Z \right]} \\
  + \proj{2} \otimes \expfn{-i \left[\phi_2 I \pm \psi_2 X\right]}.
\end{multline}
Note the inverted order of multiplication of the exponentials in the first two terms. An application of a rotary tone at an appropriate amplitude can minimize the $Y$ and $Z$ components in these products. We thereby arrive at the form of equation~\eqref{eqn:desired_wcr}.

The $\Usub{CR}$ and $\Wcr{k}$ unitaries at various stages of calibration are determined by a tomography experiment, where the target qubit is prepared in positive eigenstates of $X$, $Y$, and $Z$ operators, evolved by the gate in question, and then measured in $X$, $Y$, and $Z$ bases. Assuming the unitarity of the gate operation, the nine Pauli expectation values correspond to the entries of the three-dimensional rotation matrix of the operator, when its unitary is mapped to an element of SO(3) through the local isomorphism of SU(2) and SO(3). The unitary parameters can then be obtained through reverse mapping. This problem is in fact overconstrained since it aims to infer three real parameters from nine measurements. Therefore, in practice, the parameters are determined through a least-squares fit to the observed expectation values. Unitarity violation of the gate, i.e., qutrit decoherence or the breakdown of diagonal approximation in equation~\eqref{eqn:physical_ucr}, is detectable through this method by checking the goodness of fit. The same procedure is repeated with the control qudit prepared in $\ket{0}$, $\ket{1}$, and $\ket{2}$ states to fully reconstruct a unitary operation that is diagonal in the qutrit space. 

\subsection{\texorpdfstring{\qctwo}{Qc2} local phase for the CCZ gate}

\begin{figure*}
\centering
\begin{subfigure}[t]{0.95\linewidth}
\centering
\begin{quantikz}[column sep=0.2cm]
 & \gategroup[3,steps=12,style={dashed,rounded corners,inner sep=4pt}]{} & \wr\wr &               & \ctrl{1} &[0.3cm] \gategroup[3,steps=6,style={dashed,rounded corners,inner sep=2pt}]{} &                         &                      &               & \wr\wr &               &[0.3cm] \ctrl{1} &               & \qw \\
 & \gate{\Xplus}                                                         & \wr\wr & \gate{\Xplus} & \targ{}  & \gate{\Xplus}                                                               & \gate[2]{\Rop{xz}{\pi}} & \gate{\Rop{x}{-\pi}} & \gate{\Xplus} & \wr\wr & \gate{\Xplus} &        \targ{}  & \gate{\Xplus} & \qw \\
 &                                                                       & \wr\wr &               &          &                                                                             &                         & \gate{\Rop{z}{\pi}}  &               & \wr\wr &               &                 &               & \qw
\end{quantikz}
\caption{}\label{fig:full_ccz_sequence}
\end{subfigure}
\begin{subfigure}[t]{0.95\linewidth}
\centering
\begin{quantikz}[column sep=0.2cm]
 \lstick{$\ket{0}$} & \gate{\SX{0}} & \gate{\Usub{static}} & \gate{\Rop{z(0)}{\phi}} & \gate{\SX{0}} & \meter{} \\
\end{quantikz}
\caption{}\label{fig:ramsey_phase_measurement}
\end{subfigure}
\caption{(a) Full CCZ sequence. Two boxes represent independently basis-cycled units. (b) Ramsey-like experiment to measure the static local phase error on \qctwo. Similar to the $f_{12}$ calibration, the parameter $\phi$ is varied in small steps and the probability of observing 1 in the measurement is expressed as a function of $\phi$. The phase shift caused by $\Usub{static}$ is extracted from the resulting cosine curve.}
\end{figure*}
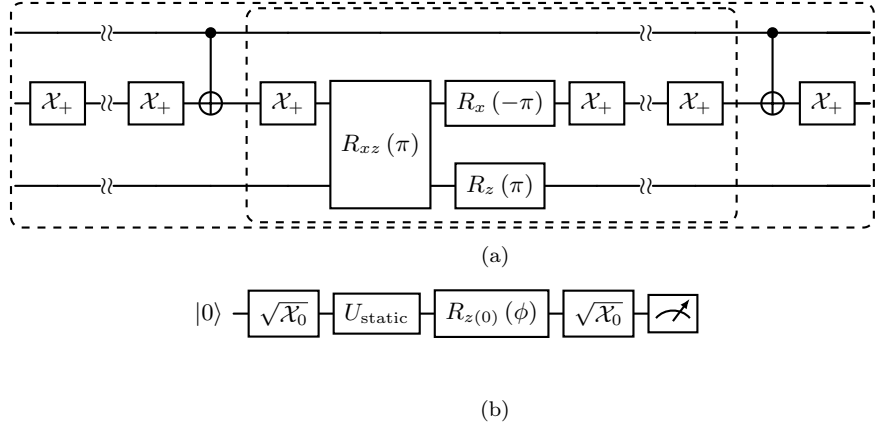

In the full CCZ sequence of Figure~\ref{fig:full_ccz_sequence}, the two boxes represent independently basis-cycled units, where dynamical phase errors from qutrit detuning are eliminated from the $\ket{0}$-$\ket{1}$ subspace but static phase errors are not. Since both boxes contain gate sequences that correspond to unitaries diagonal in the \qctwo space, the static local phase errors appear as an overall phase shift on each box. These phase shifts can be measured via Ramsey-like experiments using the circuit depicted in Figure~\ref{fig:ramsey_phase_measurement}.

%% file: error_analysis.tex
\begin{figure*}
\begin{quantikz}[column sep=0.2cm]
 \lstick{\qcone $\ket{0}$} & \gate{X}      & \gate[3]{\Usub{CCZ}} &                      &               & \meter{} \\
 \lstick{\qctwo $\ket{0}$} &               &                      &                      &               & \meter{} \\
 \lstick{\qt $\ket{0}$}    & \gate{\SX{0}} &                      & \gate{\Rop{z}{\phi}} & \gate{\SX{0}} & \meter{}
\end{quantikz}
\caption{Ramsey-like experiment circuit to measure $\Phi_{101} - \Phi_{001}$. The $R_z$ angle $\phi$ is varied in small steps, and the probability of observing 101 (qubits labeled in the order of \qt, \qcone, \qctwo) as a function of $\phi$ is fit with a cosine curve to extract the phase offset $\kappa_{101,001}$.}\label{fig:ramsey_ccz_relphase}
\end{figure*}
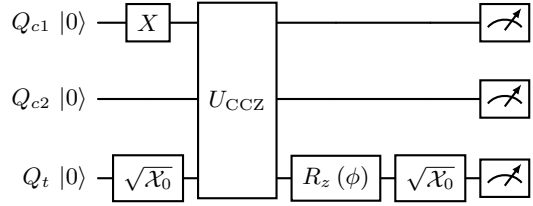

In the error analysis of the CCZ gate, $P_j$ is obtained from the truth table measurement where the initial state is prepared in eight computational basis states separately and the distribution of the final state is observed after applying the CCZ gate sequence. The phases $\Phi_j$ are measured from another Ramsey-like experiment using circuits like Figure~\ref{fig:ramsey_ccz_relphase}. Two of the three qubits are fixed to certain computational basis states, and the phase imparted by the CCZ gate onto the three-qubit system is extracted from the phase offset measurement using the remaining qubit. We repeat this experiment for each of four possible initial states on three possible qubit combinations. The resulting 12 observations correspond to phase differences between specific diagonal entries of the gate unitary. For example, with \qcone and \qctwo fixed to states $\ket{1}$ and $\ket{0}$, respectively, the measured phase difference is $\Phi_{101} - \Phi_{001}$, where the qubits are labeled in the order of \qt, \qctwo, and \qcone. We then extract seven relative phases $\Phi_j$ from the 12 observations via a least-squares fit minimizing $\sum_{j,k} (\kappa_{jk} - (\Phi_j - \Phi_k))^2 / \sigma_{jk}^2$, where $j$ and $k$ runs over measurable state label combinations, and $\kappa_{jk}$ and $\sigma_{jk}$ are the observed phase offsets and their uncertainties.

%% file: qubit_toffoli.tex
The eight-CX decomposition of the Toffoli gate in Ref.~\cite{Duckering2021-xh} is used as the reference for the qubit-based implementation. As with the main result of this paper, QPT with built-in readout error mitigation is used to measure the process fidelity. Each of the experiment circuits are repeated for 2000 shots, and circuits for multiple qubit combinations on a single device are combined and executed in parallel.

Table~\ref{tab:qubit_toffoli_fidelities} shows the full list of measured devices and qubit combinations within. The statistical uncertainty of 0.001 common to all fidelity values is omitted from the table.

\begin{table*}
\centering
\caption{List of devices and qubits used for measurements of reference gate fidelity of the qubit-only Toffoli gate decomposition. The Heron family of processors use tunable couplers, while those in the Eagle family feature static qubit couplings and CR-based entangling gates.}
\label{tab:qubit_toffoli_fidelities}
\begin{tabular}{l l c c}
\hline \hline
Device & Processor family & \qcone, \qctwo, \qt & Gate fidelity ($\pm$0.001) \\
\hline
\multirow{6}{8em}{ibm\_torino} & \multirow{6}{3em}{Heron} & 80, 92, 99 & 0.959 \\
& & 87, 88, 94 & 0.854 \\
& & 29, 30, 31 & 0.902 \\
& & 119, 120, 121 & 0.895 \\
& & 12, 18, 31 & 0.895 \\
& & 50, 51, 52 & 0.966 \\
\hline
\multirow{4}{8em}{ibm\_brisbane} & \multirow{4}{3em}{Eagle} & 4, 5, 6 & 0.862 \\
& & 52, 56, 57 & 0.895 \\
& & 8, 9, 10 & 0.796 \\
& & 115, 116, 117 & 0.869 \\
\hline
\multirow{6}{8em}{ibm\_cusco} & \multirow{6}{3em}{Eagle} & 0, 1, 2 & 0.836 \\
& & 5, 6, 7 & 0.620 \\
& & 109, 114, 115 & 0.846 \\
& & 9, 10, 11 & 0.648 \\
\hline
ibm\_kawasaki & Eagle & 83, 82, 81 & 0.775 \\
\hline
\multirow{6}{8em}{ibm\_kyoto} & \multirow{6}{3em}{Eagle} & 43, 44, 45 & 0.840\\
& & 4, 5, 6 & 0.790 \\
& & 84, 85, 86 & 0.795 \\
& & 9, 10, 11 & 0.572 \\
& & 70, 74, 89 & 0.822 \\
& & 99, 100, 110 & 0.806 \\
\hline \hline
\end{tabular}
\end{table*}